\documentclass[useAMS,usenatbib]{mnras}

\usepackage{times}
\usepackage{amssymb, amsmath}
\usepackage{graphicx,rotating,amssymb,longtable,lscape}
\usepackage{epsfig}
\usepackage{hyperref}
\usepackage{enumerate}

\def\apj{\mbox{ApJ}}
\def\apjl{\mbox{ApJL}}
\def\apjs{\mbox{ApJS}}

\def\mnras{\mbox{MNRAS}}

\def\araa{\mbox{ARA\&A}}

\def\nat{\mbox{Nature}}
\def\aap{\mbox{A\&A}}

\title[The dust covering factor in active galactic nuclei]{The dust covering factor in active galactic nuclei}
\author[M. Stalevski et al.]{Marko Stalevski$^{1,2,3,4}$\thanks{E-mail: marko.stalevski@gmail.com}, Claudio Ricci$^{4,5,6}$, Yoshihiro Ueda$^{4}$, Paulina Lira$^{1}$, Jacopo Fritz$^{7}$
\newauthor and Maarten Baes$^{3}$
\\
$^{1}$Departamento de Astronom\'\i a, Universidad de Chile, Camino El Observatorio 1515, Casilla 36-D Santiago, Chile\\
$^{2}$Astronomical Observatory, Volgina 7, 11060 Belgrade, Serbia\\
$^{3}$Sterrenkundig Observatorium, Universiteit Gent, Krijgslaan 281-S9, Gent, 9000, Belgium\\
$^{4}$Department of Astronomy, Kyoto University, Oiwake-cho, Sakyo-ku, Kyoto 606-8502, Japan\\
$^{5}$Instituto de Astrof\'\i sica, Pontificia Universidad Catolica de Chile, Casilla 306, Santiago 22, Chile\\
$^{6}$EMBIGGEN Anillo, Concepcion, Chile\\
$^{7}$Instituto de Radioastronom\'\i a y Astrof\'\i sica, IRyA, UNAM, Campus Morelia, A.P. 3-72, C.P. 58089, Mexico}

\begin{document}

\date{Accepted 2016 February 22. Received 2016 February 22; in original form 2015 November 17}

\pagerange{\pageref{firstpage}--\pageref{lastpage}} \pubyear{2016}

\maketitle

\label{firstpage}

\begin{abstract}
The primary source of emission of active galactic nuclei (AGN), the accretion disk, is surrounded by an optically and geometrically thick dusty structure (``the so-called dusty torus''). The infrared radiation emitted by the dust is nothing but a reprocessed fraction of the accretion disk emission, so the ratio of the torus to the AGN luminosity ($L_{\text{torus}}/L_{\text{AGN}}$) should correspond to the fraction of the sky obscured by dust, i.e. the covering factor. We undertook a critical investigation of the $L_{\text{torus}}/L_{\text{AGN}}$ as the dust covering factor proxy. Using state-of-the-art 3D Monte Carlo radiative transfer code, we calculated a grid of spectral energy distributions (SEDs) emitted by the clumpy two-phase dusty structure. With this grid of SEDs, we studied the relation between $L_{\text{torus}}/L_{\text{AGN}}$ and the dust covering factor for different parameters of the torus.  We found that in the case of type 1 AGNs the torus anisotropy makes $L_{\text{torus}}/L_{\text{AGN}}$ underestimate low covering factors and overestimate high covering factors. In type 2 AGNs $L_{\text{torus}}/L_{\text{AGN}}$ always underestimates covering factors. Our results provide a novel easy-to-use method to account for anisotropy and obtain correct covering factors. Using two samples from the literature, we demonstrated the importance of our result for inferring the obscured AGN fraction. We found that after the anisotropy is properly accounted for, the dust covering factors show very weak dependence on $L_{\text{AGN}}$, with values in the range of $\approx0.6-0.7$. Our results also suggest a higher fraction of obscured AGNs at high luminosities than those found by X-ray surveys, in part owing to the presence of a Compton-thick AGN population predicted by population synthesis models.
\end{abstract}

\begin{keywords}
galaxies: active -- galaxies: nuclei -- galaxies: Seyfert -- radiative transfer.
\end{keywords}

\section{Introduction}

Active galactic nuclei (AGN) are powered by accretion of gas on to a supermassive black hole \citep{Lynden-Bell1969}. The electromagnetic emission of AGN is characterised by a very strong UV/optical and X--ray continuum coming from the accretion disk and its corona. In the standard unified model of AGNs this central region is surrounded by a geometrically and optically thick dusty structure, whose exact geometry is unknown but is commonly approximated by a clumpy, toroidal distribution and referred to as ``the dusty torus'' (see reviews by \citealp{Antonucci1993} and \citealp{Netzer2015}). Part of the accretion disk continuum is radiated towards the dust--free cone cavity formed along the torus axis of symmetry and escapes freely (see Fig.~{\ref{fig:dens}). The remaining of the primary continuum goes inside the dusty region where, if the dust is optically thick, it is entirely reprocessed by scattering and absorption from dust grains, and is eventually re--emitted in the infrared (IR). Then from simple geometrical considerations it follows that the ratio of the torus to AGN luminosity ($L_{\text{torus}}/L_{\text{AGN}}$) corresponds to the fraction of the sky covered by the dust, i.e.\ the \emph{covering factor}. This covering factor is directly related to the probability to observe an AGN as a type 1 or type 2, so $L_{\text{torus}}/L_{\text{AGN}}$ lends itself naturally for the task of inferring the fraction of obscured AGNs as a function of luminosity or redshift, making it also a valuable tool for studying cosmological evolution of AGNs.

A number of works inferring the covering factor from $L_{\text{torus}}/L_{\text{AGN}}$ are present in the literature. The sample analysed by \citep{Maiolino2007} combined high-luminosity quasars and low-luminosity Type-1 AGNs for a total number of 58 objects with Spitzer--IRS data. \citet{Treister2008} estimated covering factors of 230 type 1 AGNs selected from several surveys, based on flux measured in a single Spitzer band (24 $\mu$m) and employing radiative transfer solutions. Numerous authors derived CFs by fitting torus emission models to the spectral energy distributions (SEDs) of  type 1 and type 2 AGNs \citep[e.g.][]{Fritz2006,Hatziminaoglou2009,Mor-Netzer-Elitzur2009,Mor-Trakhtenbrot2011,Alonso-Herrero2011,Ramos-Almeida2011,Mor-Netzer2012,Mateos2015,Mateos2016}. \citet{Roseboom2013} used various combinations of a hot blackbody, and a warm dusty torus, to fit broad band NIR-MIR data in a large sample of WISE-selected sources. Using broad band JHK and Spitzer mid-infrared (MIR) data, \citet{Lusso2013} analysed the obscured fraction for a sample of 513 Type 1 AGNs from the XMM-COSMOS survey. \citet{Netzer2016} use combined Herschel/SPIRE and WISE observations of 35 high-redshift AGNs. Most of these works have broadly consistent results, finding general trends of decreasing obscured fraction with increasing $L_{\text{AGN}}$ \citep[see the discussion and comparison in][]{Lusso2013}. The exception is the work by \citet{Netzer2016} who infer covering factors consistent with no clear evolution with nuclear luminosity, within the uncertainties for the bolometric correction factor.

However, the above reasoning for interpreting $L_{\text{torus}}/L_{\text{AGN}}$ as the covering factor is strictly valid only if both the disk and the surrounding dusty structure are emitting isotropically. The standard geometrically-thin and optically thick disk cannot emit isotropically \citep{Netzer1987}. To obscure the central region, as required by the unification model, the dusty torus has to be optically thick. Furthermore, it has to be optically thick to its own radiation in the MIR, to produce the observed appearance of the silicate feature in emission and absorption (in the first case it can often be weak or flat) in type 1 and 2 AGNs. As such, the dusty torus will inevitably emit anisotropically as well. This means that the observed $L_{\text{torus}}/L_{\text{AGN}}$ will not represent the actual covering factor, but it will rather be convolved with the aforementioned effects due to the anisotropic nature of the accretion disk and the dusty torus emission. Disentangling these effects to reveal the true covering factor cannot be done analytically. Instead, authors usually discuss their results within the limiting cases of total isotropy and total anisotropy \citep[e.g.][]{Lusso2013,Netzer2015}. The disadvantage of this approach is that these limits represent two extremes between which considerable uncertainty remains. 

Given the complex nature of the problem, the most promising way to properly account for these effects is to realistically simulate reprocessing of the accretion disk radiation as it travels through the dusty medium and reconstruct the resulting IR emission. This is exactly what is our approach in the present work. We employ a state--of--the--art radiative transfer code based on the Monte Carlo technique (MCRT) to calculate a grid of SEDs for an inhomogeneous dusty torus for different values of covering factor and other parameters. This grid of SEDs allow us to study the relation between $L_{\text{torus}}/L_{\text{AGN}}$ and the covering factor. We then use the obtained relations to correct the observed $L_{\text{torus}}/L_{\text{AGN}}$ and recover the actual covering factors. A somewhat similar approach was adopted by \citet{Treister2008} who used radiative transfer emission models to relate the observed ratio of 24~$\umu$m and bolometric luminosities to the obscured fraction of AGN sky. However, the realization of our model is substantially different, we cover a larger parameter space in our simulations, and we focus on the deeper understanding of the relation between $L_{\text{torus}}/L_{\text{AGN}}$ and the dust covering factor, by isolating the different effects that contribute to the torus anisotropy.

In Sec.~{\ref{sec:mod}} we overview our method and provide the details of our dusty torus model. We present the relevant properties of the calculated grid of model SEDs in Sec.~{\ref{sec:res}} and examine in detail the obtained relations between $L_{\text{torus}}/L_{\text{AGN}}$ and the covering factor. In Sec.~{\ref{sec:discuss}} we apply our method on two samples from the literature and discuss implications for inferring the obscured fraction of AGNs as a function of luminosity. The paper is concluded with a summary of our main findings in Sec.~{\ref{sec:con}}.

\section{Method and model}
\label{sec:mod}

\subsection{Method}

A considerable source of uncertainties in inferring the dust covering factor from $L_{\text{torus}}/L_{\text{AGN}}$ comes from the bolometric correction factor which is needed to estimate $L_{\text{AGN}}$. According to \citet{Marconi2004}, this correction factor for luminosity in optical regime decreases with increasing $L_{\text{AGN}}$. However, the correction factors adopted by different authors differ significantly \citep[e.g.][]{Shen2011,Trakhtenbrot-Netzer2012,Runnoe2012,Krawczyk2013}. In fact, \citet{Netzer2016} argue that earlier results of IR-derived covering factors are biased by the inconsistent use of various bolometric correction factors. This issues is beyond the scope of the present work; we will assume that $L_{\text{AGN}}$ can be constrained with satisfactory accuracy and focus only on $L_{\text{torus}}$ and uncertainties due to anisotropic emission.

Our approach to the problem is the following. We calculate a grid of IR SEDs emitted by the dusty torus for a given central source luminosity ($L_{\text{AGN}}$) with anisotropic emission pattern. Then, we integrate the IR flux of the model SEDs, convert it to luminosity, and compare $L_{\text{torus}}/L_{\text{AGN}}$ to the true covering factor, which in our model corresponds to the opening angle of the torus. By means of this comparison we investigate how the $L_{\text{torus}}/L_{\text{AGN}}$ ratio deviates from the actual covering factor, and how this deviation depends on different parameters of the obscuring dust. Apart from the total infrared emission, we also test near-infrared (NIR), MIR, far-infrared (FIR) and individual fluxes at 6.7~$\umu$m and 12~$\umu$m as proxies of the dust covering factor \citep[e.g.][]{Maiolino2007}.

\subsection{The \textsc{SKIRT} code}

To calculate model SEDs we make use of \textsc{SKIRT}\footnote{\url{http://www.skirt.ugent.be}}, a state--of--the--art code for simulating continuum radiation transfer in dusty astrophysical systems \citep{Baes2003,Baes2011,Camps-Baes2015}. \textsc{SKIRT} employs the Monte Carlo technique to emulate the relevant physical processes including multiple anisotropic scattering, absorption and (re-)emission by the dust, and handles any 3D geometry without limitation \citep{Saftly2013,Saftly2014,Camps2013}. The code is being used for a variety of applications, including study of the dust energy balance in spiral galaxies \citep{Baes2010,DeLooze2012a,DeLooze2012b,DeGeyter2015}, and the investigation of the structure and observable properties of AGN dusty tori \citep{Stalevski2012a,Stalevski2012b,Stalevski2012c,Popovic2012}.

\subsection{The dusty torus as a two-phase medium}

We model the distribution of the dust in the torus as a two-phase medium, consisting of a large number of high-density clumps embedded in a smooth dusty component of low density. Previously, we have found that an advantageous property of such two-phase medium is that it can produce attenuated silicate features, while at the same time have a pronounced NIR emission, which is challenging for both smooth and clumpy models alone \citep{Stalevski2012a,Stalevski2012c}. Since then, several lines of evidences were found providing further supporting that the torus is indeed a multiphase structure. Studies using hydrodynamical simulations and taking into account other processes such as self--gravity of the gas, radiative cooling and heating due to supernovae or accretion disk, found that the interstellar medium around the AGN would result in a multiphase filamentary structure \citep{Wada2009,Wada2012}. Using the two--phase model SEDs grid of \citet{Stalevski2012a}, \citet{Roseboom2013} were able to reproduce the observed ratio of NIR to total IR luminosity in a large sample of sources selected from the WISE catalogue, something which models consisting of only clumps are not able to achieve. Analysing a large AGN sample from the WISE catalogue, \citet{Assef2013} found that the observed reddening distribution can be explained by a torus-like structure in which thick dust clouds are embedded in a diffuse inter-cloud dust medium. Another indication of a presence of smooth, low-density inter-cloud medium comes from constant baseline level absorption, between the strong absorption events, in type 2 AGN well-monitored with RXTE \citep{Markowitz2014}. Studying variable reddening in Narrow-line Seyfert 1, \citet{Leighly2015} concluded that it is ``plausible that the occulting material is a clump embedded in a larger region of dusty gas responsible for the longer timescale reddening changes, and both are associated with the torus''. Finally, such two-phase structures have been actually observed in the central regions of Milky Way (the so-called Central Molecular Zone and Circumnuclear Disk) and it has been suggested that they represent a remnant of a dusty torus that may have played a role in past AGN phases of our Galaxy \citep{Molinari2011,Ponti2013}.

\subsection{Model parameters}
\label{sec:mod-par}

The dusty torus model from this work shares a common ground with the one presented in \citet{Stalevski2012a}, but with a number of changes and improvements. Here we describe the most important properties.

The primary source of emission is the accretion disk, approximated by a central point source with anisotropic emission distributed as proposed by \citet{Netzer1987}:
\begin{equation}\label{eqn:anisodisk}
L(\theta) \propto \cos\theta(2\cos\theta+1) ,
\end{equation}
where $\theta$ is polar angle of the coordinate system. The first factor in this equation describes the change in the projected surface area and the second factor accounts for the limb darkening effect.
The SED of the accretion disk is described by the following composition of power laws, in agreement with theoretical predictions and observational evidences \citep[e.g][]{Hubeny2001,DavisLaor2011,SloneNetzer2012,Capellupo2015}:
\begin{equation}\label{eqn:disksed}
\lambda L_{\lambda}\propto\left\{
\begin{array}{lrrr}
\lambda^{1.2}  &  \; 0.001 \leq \lambda < \leq 0.01 & [\mu\text{m}]\\
\lambda^{0}    &  \; 0.01  < \lambda \leq 0.1   & [\mu\text{m}]\\
\lambda^{-0.5} &  \; 0.1   < \lambda \leq 5     & [\mu\text{m}]\\
\lambda^{-3}   &  \; 5     < \lambda \leq 50    & [\mu\text{m}]
\end{array}
\right.
\end{equation}
We approximate the spatial distribution of dust with a flared disk whose geometry is defined by the inner ($R_{\text{in}}$) and outer radii ($R_{\text{out}}$) and the half opening angle ($\Delta$). The inner radius is determined by the dust sublimation temperature for a given $L_{\text{AGN}}$ \citep{Barvainis1987}:
\begin{equation}\label{eqn:rin}
  \left(\frac{R_{\text{in}}}{\text{pc}}\right)
  \simeq
  1.3
  \left(\frac{L_{\text{AGN}}}{10^{46}~{\text{erg}}\,{\text{s}}^{-1}}\right)^{0.5}
  \left(\frac{T_{\text{sub}}}{1500~{\text{K}}}\right)^{-2.8} ,
\end{equation}
assuming an average dust grain size of 0.05~$\umu$m. The accretion disk emission described by Eq.~(\ref{eqn:anisodisk}) is highly anisotropic, with the strongest emission perpendicular to the disk and none in the equatorial plane. The inner radius in this case cannot be constant but must follow the same dependence with polar angle:
\begin{equation}\label{eqn:anisorad}
R_{\text{in}} \propto R_{\text{iso}}[\cos\theta(2\cos\theta+1)]^{0.5} ,
\end{equation}
where $R_{\text{iso}}$ is inner radius in the case of isotropic disk emission.

In the literature, some authors use the term ``half opening angle'' to describe the opening of the dust-free cone measured from the polar axis, while other authors use it as a measure of the dust-filled zone, from the equator to the edge of the torus. To remain consistent with the notation used in our original model in \citet{Stalevski2012a}, we choose the latter definition, measuring the half opening angle from the equator to the flaring edge. This parameter is linked to the covering factor (CF) by a simple relation:

\begin{equation}\label{eqn:cfgeo}
\text{CF} = \sin(\Delta) .
\end{equation}

Dust is distributed according to a law that allows a density gradient along the radial direction ($r$) and with polar angle ($\theta$):
\begin{equation}\label{eqn:dustdens}
\rho\left(r,\theta \right)\propto r^{-p}e^{-q|\cos\theta|} ,
\end{equation}
where $r$ and $\theta$ are coordinates in the adopted coordinate system. Dust in our model is modeled as a mixture of silicate and graphite grains with classical MRN size distribution \citep*{MRN1977}, with the normalization factors for size distribution from \citet{Weingartner-Draine2001}, and with optical properties taken from \citet{Laor-Draine1993} and \citet{Li-Draine2001}. 

The parameters defining the clumpiness are the total number of clumps ($N_{\text{cl}}$) and the fraction of total dust mass of the torus locked up inside clumps ($f_{\text{cl}}$). The clumps are allowed to overlap and form complex structures. For density profile of individual clumps we assume the standard smoothing kernel with compact support, providing computational accuracy and efficiency \citep{Monaghan-Lattanzio1985,Springel2010}:
\begin{equation}\label{eqn:cubsplie}
W(u) =\frac{8}{\pi} \left\{
\begin{array}{ll}
1-6 u^2 + 6 u^3, &
0\le  u \le\frac{1}{2} ,\\
2\left(1-u\right)^3, & \frac{1}{2}< u \le 1 ,\\
0 , & u>1 ,
\end{array}
\right.
\end{equation}
with $u$ being the radius normalised by the scale radius of a clump, $h$. Fig.~{\ref{fig:dens}} illustrates the geometry and dust distribution in a typical two-phase model. For details on the implementation of such a clumpy geometry see \citet{Baes-Camps2015}.
\begin{figure*}
\centering
\includegraphics[height=0.49\textwidth]{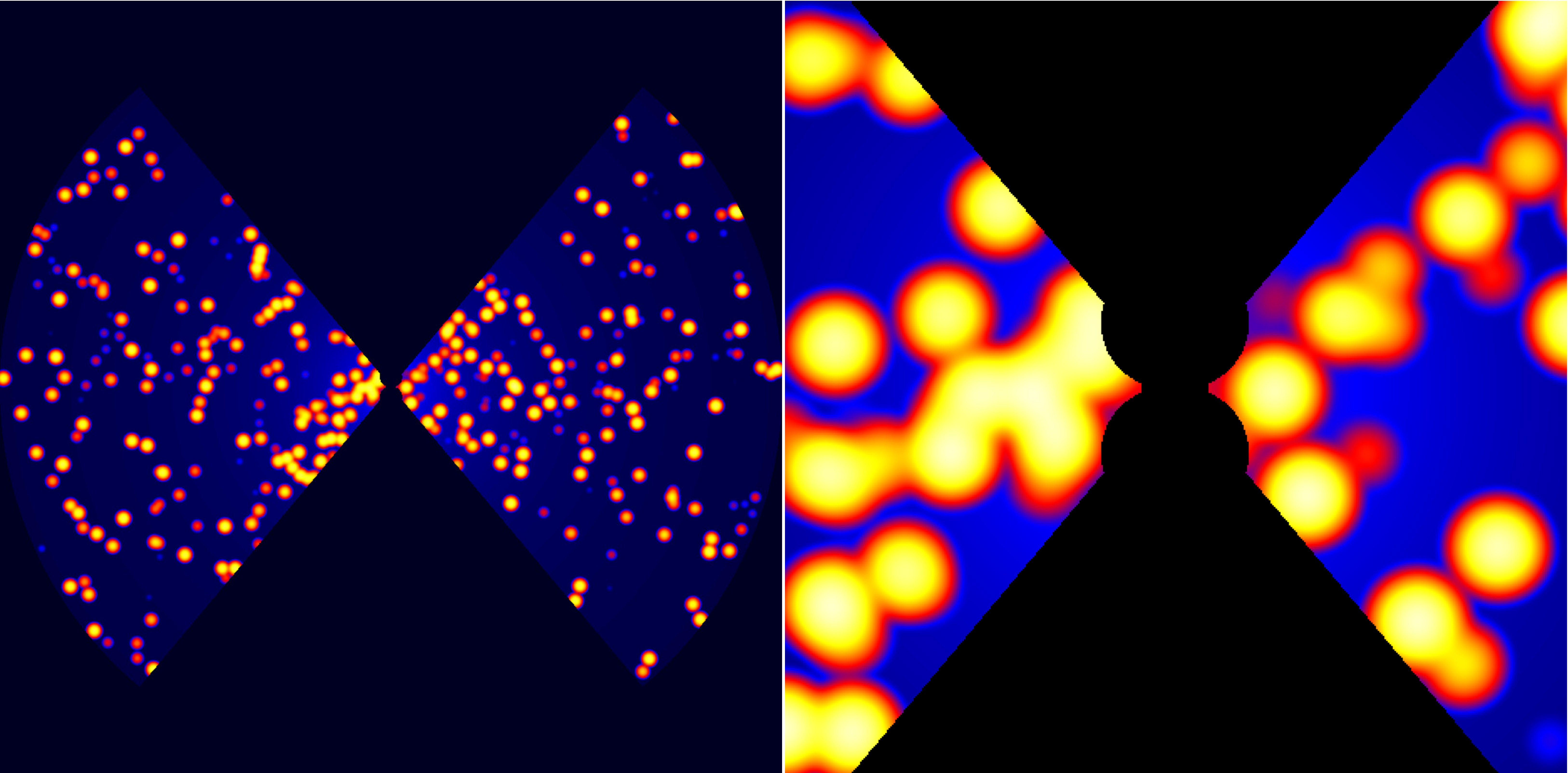}
\caption{Dust density map of the vertical xz plane (left) and zoom-in (right) of the inner region revealing the inner wall reshaped to account for anisotropic irradiation by the accretion disk. High-denisity clumps are seen in yellow, while the low-density interclump dust is in shades of blue. Presented in logarithmic color scale.}
\label{fig:dens}
\end{figure*}

The total amount of dust is set by the equatorial optical depth at 9.7~$\umu$m ($\tau_{9.7}$) of the underlying density profile defined by Eq.~(\ref{eqn:dustdens}), before applying the algorithm that generates clumps. After redistributing the dust into a clumpy two-phase medium, the optical depth along a given line of sight can vary significantly, mostly depending on the number of clumps intercepted, as seen in the optical depth map shown in Fig.~{\ref{fig:tau}}. In this map, dark blue areas represent clump-free lines of sight. The adopted values of the parameters defining clumpiness result in the contrast (i.e. ratio) between the high and low dust density phases of $\approx 100$. However, even for such a high contrast, the clump-free lines of sight are still optically thick for the accretion disk radiation ($\tau_{V}=5-10$), so the covering factor still corresponds to the torus opening angle (Eq.~\ref{eqn:cfgeo}).

\begin{figure*}
\centering
\includegraphics[height=0.36\textwidth]{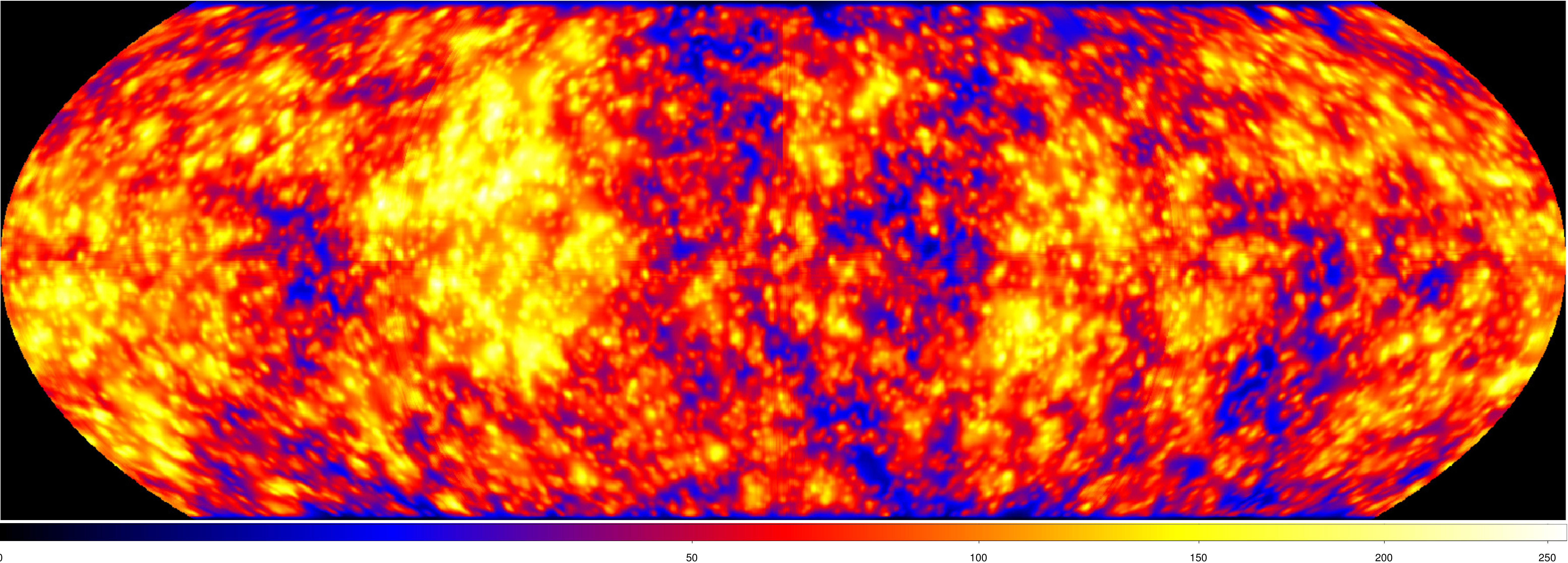}
\caption{Optical depth map of the sky in the V-band, as it would be seen from the center of AGN for the density profile show in Fig.~{\ref{fig:dens}}. The dark blue areas are the ones with the lowest optical depth $\tau_V\approx 5-10$. Presented in square root color scale.}
\label{fig:tau}
\end{figure*}

\section{Results and discussion}
\label{sec:res}

In this section we first present examples of typical model SEDs and images and comment on some of their properties relevant for this work. Then we present and analyse the relation between $L_{\text{torus}}/L_{\text{AGN}}$ and the dust covering factor.

\subsection{Model SEDs and images}
\label{sec:sed}

In Figs.~{\ref{fig:sed-oa}} and {\ref{fig:sed-tau}} we present SEDs with typical model parameters adopted for this work. The luminosity of the primary source (the accretion disk) is $L_{\text{AGN}}=10^{11} L_{\odot}$ and the SEDs in all the figures are scaled to a source at a distance of 10~Mpc. We remind the reader of some important scaling properties of dust radiative transfer models: when a change of the primary source luminosity is accompanied by a required change of the sublimation radius (Eq.~\ref{eqn:rin}), while keeping the optical depth and the outer--to--inner radius ratio ($R_{\text{out}}/R_{\text{in}}$) constant, the shape of the resulting SED will remain exactly the same, only scaled along the flux axis \citep{Ivezic-Elitzur1997,Fritz2006}. Also, it is trivial to rescale the model flux to any given distance between the observer and the source. Thus, model SEDs obtained with MCRT method are applicable to any given source luminosity and distance (see \citealp{Honig-Kishimoto2010} for more details).

Graphite and silicate grains have different sublimation temperatures, with graphite being able to withhold higher ones and thus survive closer in. Also, larger grains are cooling more efficiently than smaller grains. Hence, rather than a well defined $R_{\text{in}}$ determined by the sublimation radius, we expect to have a sublimation zone, from larger to smaller grains and from graphite to silicate (see e.g., \citealp{Kishimoto2007,Mor-Netzer2012}). However, setting up the sublimation zone is non-trivial, as radiative transfer and shielding effects between the different zones must be taken into account and the simulation has to be iterated until each zone reaches its sublimation temperature \citep{Fritz2006}. A proper treatment of sublimation zone is beyond the scope this work and will be part of a future investigation. For this work we choose the same inner radius for both graphite and silicates corresponding to an average temperature of $\approx 1250$~K of the regions located closest the central source ($R_{\text{iso}}=0.21$~pc; the actual $R_{\text{in}}$ is set by Eq.~(\ref{eqn:anisorad})) As a consequence, this may result in lower NIR emission in our models than in some observed composite SEDs, such as those found in \citet{Mor-Netzer2012}. On the other hand, \citet{Roseboom2013} found that the observed ratio of NIR to total IR luminosity in their WISE-selected sample is easily achievable by our models presented in \citet{Stalevski2012a}. In any case, we base our main results on the total IR luminosity as the covering factor proxy. The eventual lack of hot dust may results in redistribution of flux between NIR, MIR and FIR, but the total torus luminosity would remain the same; we remind the reader that the total luminosity radiated by the torus is completely determined by the fraction of accretion disk radiation that penetrates the torus through the dusty interface. 

Finally, in the examples shown here, the outer torus radius is 5~pc, the dust density law (Eq.~\ref{eqn:dustdens}) is taken to have an exponent $p=1$ for the radial part and to be constant with polar angle ($q=0$), and the edge-on optical depth ($\tau_{9.7}$) takes values between 0.1 and 10. Scale radius of a clump ($h$) is 0.4~pc. Only $\approx3\%$ of the total dust mass is in the low-density phase ($f_{\text{cl}}\approx0.97$). 

The full line in Fig.~{\ref{fig:sed-oa}} represents the accretion disk SED; the remaining lines are the SEDs of our torus model for different half opening angles in the range of $\Delta=10-80^\circ$. The model SEDs have the general properties consistent with those previously presented in \citet{Stalevski2012a}. It is worth inspecting closer the SEDs of models with low covering factors ($\Delta=10,20^\circ$). A disk with an anisotropic luminosity as described by Eq.~{\ref{eqn:anisodisk}} emits very little in the directions close to equatorial plane. The tori with small covering factors thus receive very little illumination and as a consequence have very weak IR dust signatures. In fact, IR-based selection would most likely completely miss these objects.

In Fig.~{\ref{fig:sed-tau}} we show how the SEDs change as a function of the optical depth, from very low to high. Here we draw the attention of the reader to the three cases of low optical depths in the MIR, $\tau_{9.7} = 0.1, 0.5, 1$. We see that these three cases result in strong 10~$\umu$m silicate emission feature even in edge-on view. As type 2 AGNs, with a few exceptions, show this feature in absorption, we consider these cases to be unrealistic and include them only for the sake of completeness, and to isolate the effects of the accretion disk and dust anisotropy.

We also present the examples of model images at different wavelengths and temperature maps. The first panel from the left in Fig.~{\ref{fig:img}} displays the V-band image where scattered light is dominant. The remaining panels display images of hot and warm dust at 4.6, 9.6 and 12~$\umu$m. Fig.~{\ref{fig:temp}} shows the temperature distribution in the $xz$ (left) and $yz$ (right) planes. The patterns seen in these maps are formed by alternation of clump-free lines of light propagation and regions shadowed by the clumps.

\begin{figure}
\centering
\includegraphics[height=0.34\textwidth]{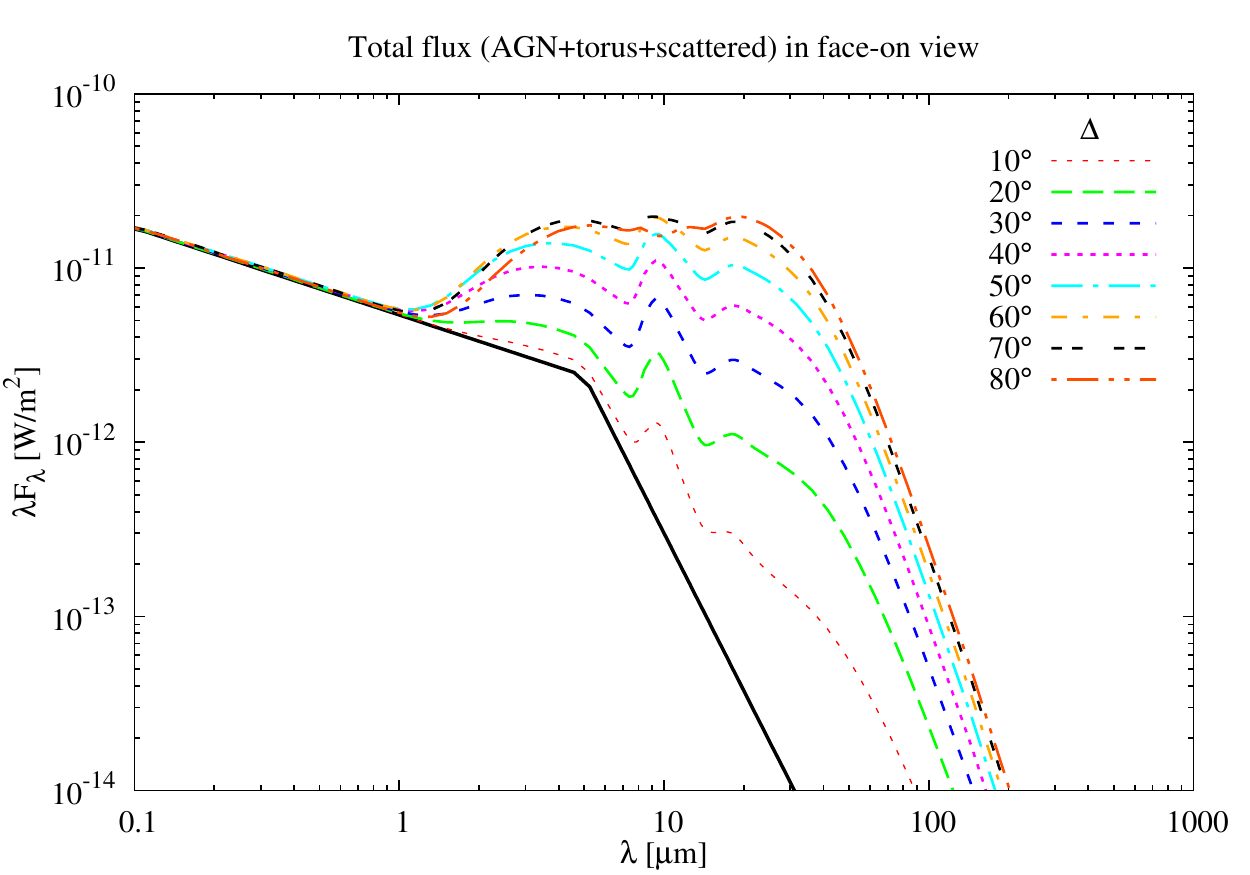}
\caption{SEDs for different values of the torus half opening angle $\Delta$ (i.e. the covering factor). The solid line is the central source (accretion disk) SED.}
\label{fig:sed-oa}
\end{figure}
\begin{figure*}
\centering
\includegraphics[height=0.34\textwidth]{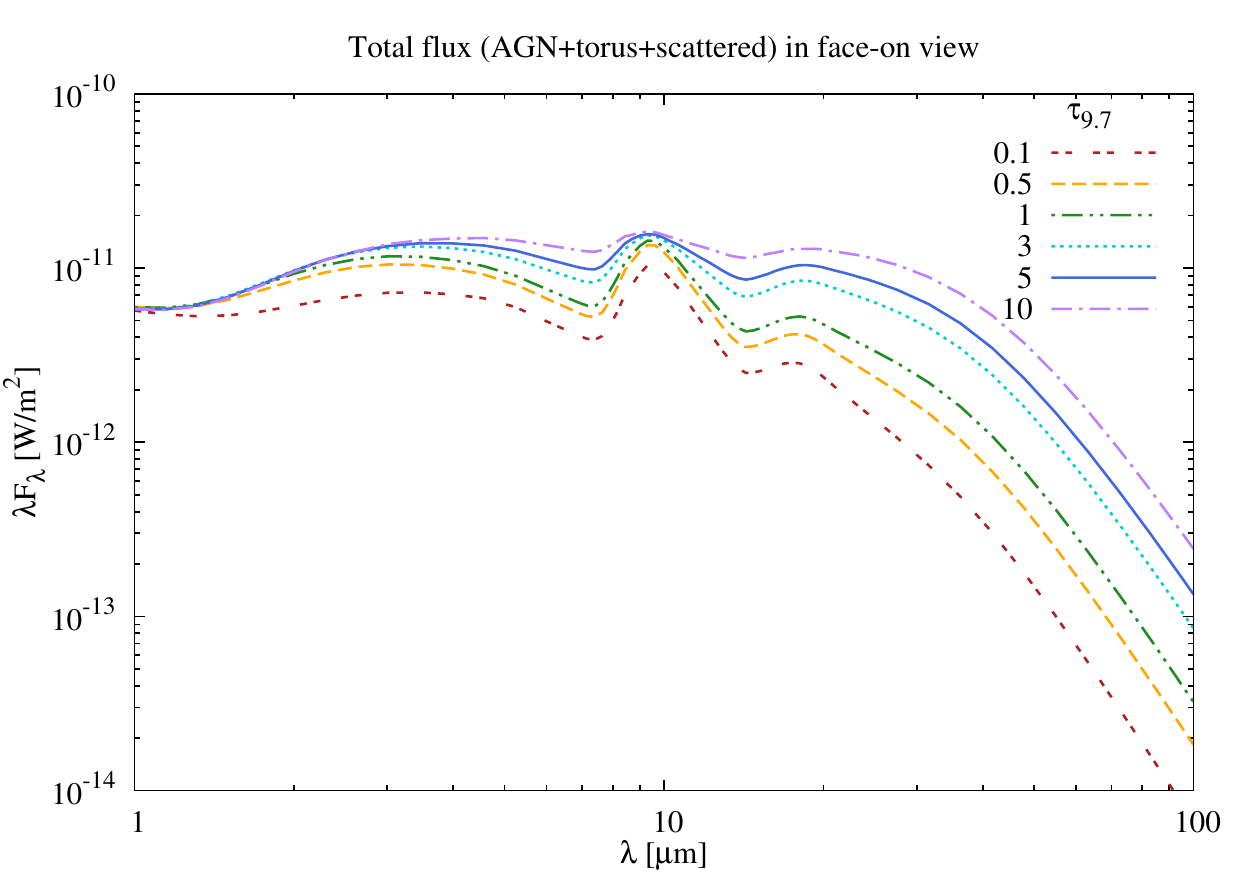}
\includegraphics[height=0.34\textwidth]{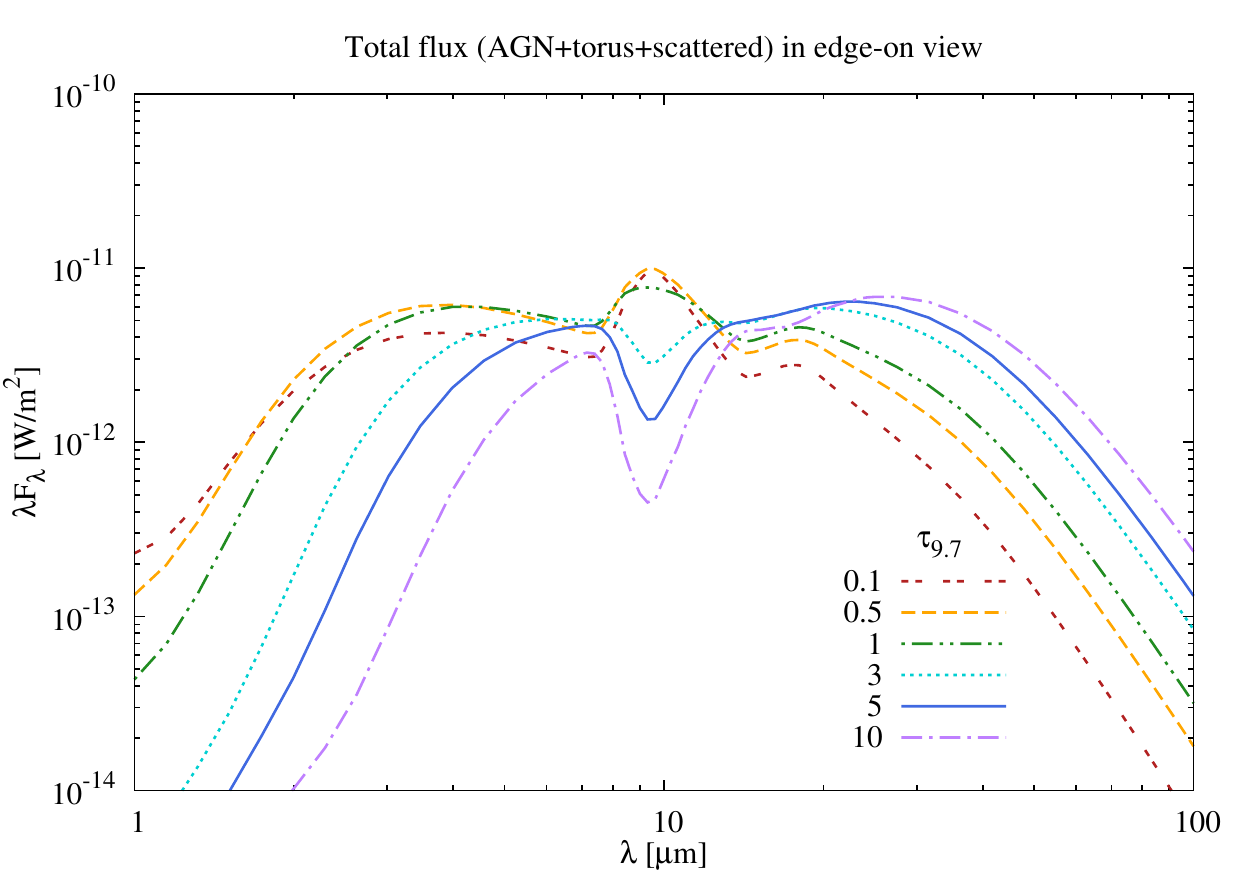}
\caption{Illustrating the change of the model SEDs for different optical depths along the equatorial line of sight ($\tau_{9.7}$) for face-on (left) and edge-on (right) views. The half opening angle is fixed at $\Delta=50^{\circ}$. Note how the models with low optical depth (0.1--1) result in strong 10~$\umu$m silicate emission feature even in edge--on views.}
\label{fig:sed-tau}
\end{figure*}
\begin{figure*}
\centering
\includegraphics[height=0.255\textwidth]{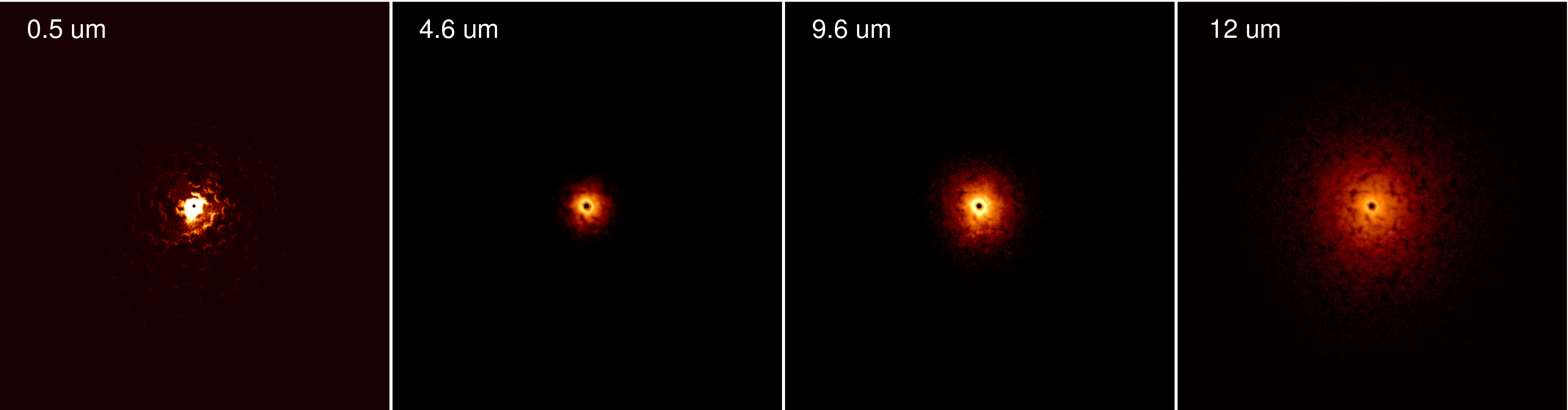}
\caption{Flux images of the torus in face-on view at different wavelengths, from the optical to MIR. In the left panel we see scattered light in optical; continuing to the right, the other panels show hot to warm dust emission, seen in the NIR-MIR regimes. Presented in logarithmic color scale with arbitrary cut offs in each image.}
\label{fig:img}
\end{figure*}
\begin{figure*}
\centering
\includegraphics[height=0.53\textwidth]{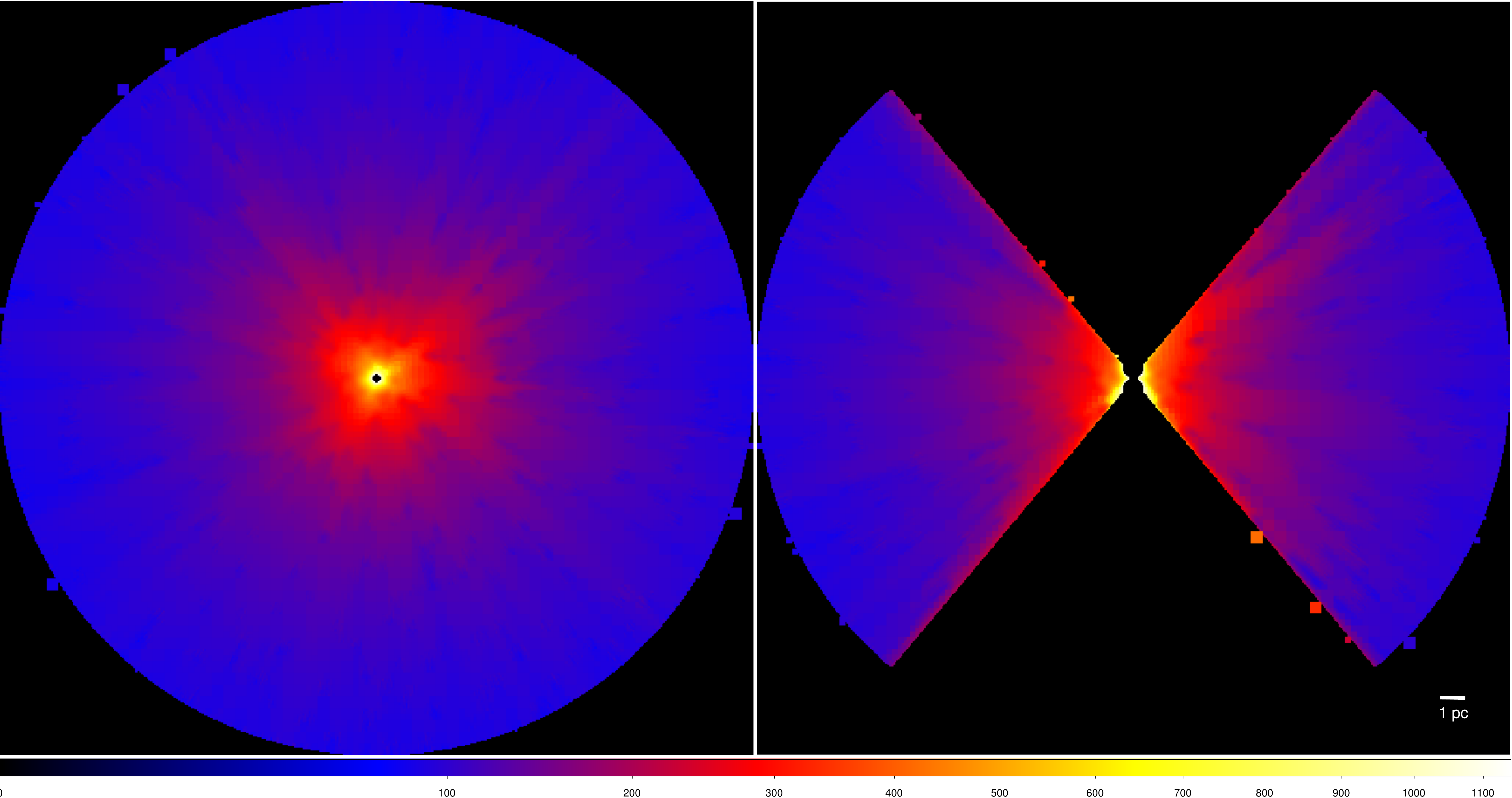}
\caption{Temperature maps of the $xz$ (left) and $yz$ (right) planes in the Kelvin scale. Presented in square root color scale. The alternation of clump-free lines of light propagation and regions in shadow behind the clumps is visible. White line in the lower right corner represents 1~pc, which is equivalent to $\approx 2 R_{\text{sub}}$ for a given luminosity, $L_{\text{AGN}}=10^{11} L_{\odot}$.}
\label{fig:temp}
\end{figure*}

\subsection{$L_{\text{torus}}/L_{\text{AGN}}$ as a covering factor proxy: type 1 AGN case}
\label{sec:cf-type1}

Now we will investigate the relation between $L_{\text{torus}}/L_{\text{AGN}}$ and the actual covering factor of the dust, and its dependence on the different parameters of the torus, focusing on the case of type 1 AGNs. All the results presented in this section are obtained for face-on viewing angle {$i=0^\circ$}. However, the torus IR SEDs change only minimally with changing viewing angle, as long as it still provides a dust-free (type 1) line of sight \cite{Stalevski2012a}. Thus, the results presented here are valid for type 1 AGNs in general irrespective of their inclination. 

\subsubsection{The total infrared emission}

\begin{figure*}
\centering
\includegraphics[height=0.49\textwidth]{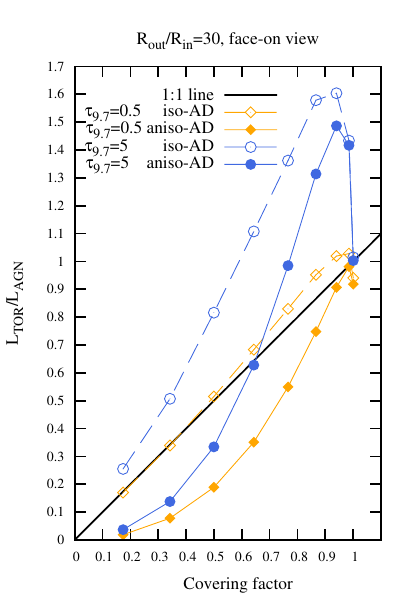}
\includegraphics[height=0.49\textwidth]{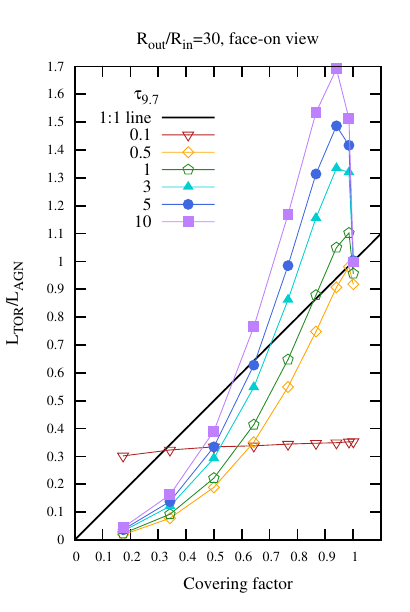}
\includegraphics[height=0.49\textwidth]{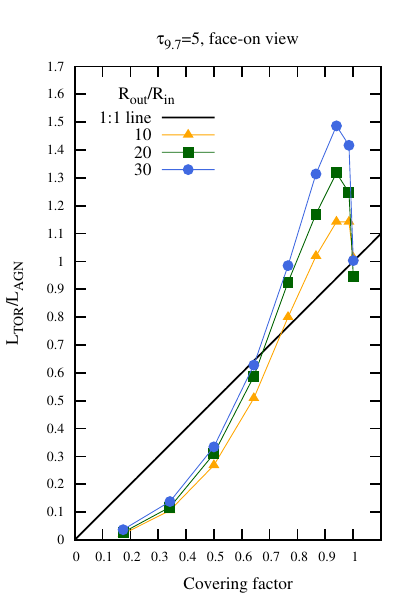}
\caption{The relation between $L_{\text{torus}}/L_{\text{AGN}}$ and the covering factor for the case of face-on view (type 1 AGN). The left panel shows cases of iso- and anisotropic disk emission (empty and full symbols) and optically thin and thick dusty tori (dashed and solid lines). Full circles with blue lines represent a realistic case of an anisotropic disk and MIR thick dust. The central panel shows the same relation for a range of $\tau_{9.7}$ and with an anisotropic disk in all the cases. Full symbols denote realistic cases with the dusty torus being moderately to highly MIR thick. Right panel: the same as the central panel but for different values of the $R_{\text{out}}/R_{\text{in}}$ parameter. The main aspect to note is the that, in the realistic case (e.g. central panel, full symbols), $L_{\text{torus}}/L_{\text{AGN}}$ underestimates low covering factors and overestimates high covering factors.}
\label{fig:cf-type1}
\end{figure*}

We start by examining Fig.~{\ref{fig:cf-type1}}. The horizontal axis in this plot denotes the covering factor (defined by the parameter $\Delta$, the half opening angle) while on the vertical axis we plot the $L_{\text{torus}}/L_{\text{AGN}}$ ratio. Both $\Delta$ and $L_{\text{AGN}}$ are input parameters in the model. $L_{\text{torus}}$ is measured from the model SED by simply integrating the dust flux over the entire wavelength range and converting it to luminosity (by multiplying it by $4\pi$ and the square of the distance to the source). The integration is done over the entire wavelength range of the simulation, but the same results are obtained if the integration is limited to the range in which dust emission is significant, 1--100~$\umu$m. To facilitate a visual inspection, we include the identity 1:1 line in the plots. If the $L_{\text{torus}}/L_{\text{AGN}}$ ratio is a perfect covering factor proxy, all the points would be found exactly on this line; the further away they are from the line, the less reliable covering factor proxy it is. The parameters of the models used in this figure correspond to those described in section \ref{sec:sed}. Through out the text, the expression ``optically thin/thick'' refers specifically to the V-band, while ``MIR thin/thick'' refers to 9.7~$\umu$m.

Let us first focus on the left panel of Fig.~{\ref{fig:cf-type1}}, where we explore the effects of iso- and anisotropic disk emission and optical thickness of the obscuring dust. Empty symbols with dashed lines correspond to the isotropic disk emission, while full symbols with solid lines to the anisotropic case. Yellow lines correspond to optically (and MIR) thin tori, while blue lines to MIR-thick. In the case of isotropically emitting disk surrounded by MIR thin dust, we expect the torus emission to be almost perfectly isotropic and $L_{\text{torus}}/L_{\text{AGN}}$ to be a very good proxy of the covering factor. As we see from the plot this is indeed the case (empty diamonds with dashed yellow line). 

Still considering isotropic disk emission, let us see what happens if the dust surrounding it is MIR thick (empty circles with dashed blue line). The torus is now optically thick to its own radiation, resulting in an overall anisotropic emission pattern in which type 1 inclinations have larger IR luminosity than type 2. As a consequence, covering factors assessed by $L_{\text{torus}}/L_{\text{AGN}}$ are overestimated. 

Next, we consider the case of anisotropic disk emission (Eq.~{\ref{eqn:anisodisk}}) with MIR thin dust (full diamonds with solid yellow line). As mentioned earlier, the reasoning that allows us to interpret the $L_{\text{torus}}/L_{\text{AGN}}$ as the dust covering factor is strictly valid only if both the disk and the torus are emitting isotropically. Since we are considering MIR optically thin dust, the torus is indeed emitting isotropically. But since that is not the case with the accretion disk, $L_{\text{torus}}/L_{\text{AGN}}$ is no longer a valid proxy of the covering factor: we see that in this case the luminosity ratio underestimates it. 

Finally, we examine the realistic case of anisotropic disk emission and MIR thick dust (full circles with blue solid line). This time we expect to have a combination of the two previous, simplified cases. We see this is indeed the case: we recover the correct covering factor at $\approx0.65$, while the lower values are underestimated and higher values are overestimated. We notice that $L_{\text{torus}}/L_{\text{AGN}}$ significantly deviates from the covering factor and that the difference is not just a systematic shift: the deviation strongly depends on the covering factor itself. Another undesirable characteristic is the non--uniqueness of the relation in the ${\text{CF}}=0.75-1$ range (i.e. there is no one-to-one relation). As the covering factor is increasing and the dusty structure starts to resemble a sphere, the IR emission becomes more isotropic. Thus the luminosity ratio starts to approach the 1:1 line. The rightmost point in the plot is the case of a model with a covering factor of 1, i.e.\ a sphere of dust fully obscuring the central engine. A dusty sphere is indeed emitting isotropically and we recover the correct value of the covering factor. However, the turning point is at ${\text{CF}}\approx0.94$; for such high covering factor there is a very low probability that the objects would be observed as type 1, so this issue of non-uniqueness will not have a significant impact on actual measurements.

We will now focus on the central panel of Fig.~{\ref{fig:cf-type1}}. From now on, we are considering only realistic cases of anisotropic disk emission. Here, we look at how the relation between $L_{\text{torus}}/L_{\text{AGN}}$ and the covering factor changes over a wide range of dust optical depths, $\tau_{9.7} = 0.1-10$. In the plot, empty symbols denote MIR thin cases, full symbols MIR thick cases. For $\tau_{9.7}= 0.1$ (red empty triangles) the dust is so optically thin that a significant fraction of the AGN emission is able to pass through the torus without being reprocessed by the dust. In this case, the assumption that $L_{\text{torus}}/L_{\text{AGN}}$ may be used as a covering factor proxy completely breaks down. Besides, such an optically thin torus cannot provide obscuration required by the unified AGN model.

The following two values of the optical depth we have in our grid ($\tau_{9.7}=0.5, 1$) correspond to a dusty region that is opaque to the disk emission. However, as we showed in Fig.~{\ref{fig:sed-tau}}, in these cases the torus is emitting isotropically, resulting in a strong 10~$\umu$m silicate emission feature even for edge-on views. This is in contradiction with observations of type 2s, which, with a few exceptions, show the feature in absorption. Thus, we will be concentrating on the cases of moderate to high optical depths, $\tau_{9.7}= 3, 5, 10$. Focusing only on these cases, we notice a favourable characteristic: the relation between $L_{\text{torus}}/L_{\text{AGN}}$ and covering factor for different optical depths differ significantly from each other only at the higher end of the covering factors.

Now we examine the right panel in Fig.~{\ref{fig:cf-type1}}, where we see how $L_{\text{torus}}/L_{\text{AGN}}$ behaves for different values of $R_{\text{out}}/R_{\text{in}}$ parameter (10, 20, 30; from compact to extended tori). Again, we consider only anisotropic disk emission and moderate optical thickness of the torus ($\tau_{9.7}=5$). We see that the luminosity reacts to the changes of this parameter in a similar manner as in the previously considered case of different optical depths. 

We could demonstrate the same exercise for other torus parameters, but this is not necessary. The exact shape of SED is difficult to predict as it results from the combined effects of all the parameters. However, for this work only the total torus luminosity is relevant, and the level of its anisotropy. The exact level of anisotropy will be determined by an intricate combination of the effects that different parameters have on the radiation transfer. However, the parameter that has the strongest and most direct influence on the anisotropy is the optical depth of the dust. This leads us to a very important point: \emph{all possible combinations of effects other parameters have on the level of anisotropy will be within the limits of the cases of optically thin-to-thick dusty tori.} And we saw that in the realistic range of moderate-to-high optical depth of the dust in the MIR, the different cases diverge significantly from each other only at the high end of the covering factor values. Thus, we can use the obtained relations to correct the observed $L_{\text{torus}}/L_{\text{AGN}}$ ratio so that it corresponds more closely to the actual covering factor. While we may not know the exact optical depth of the objects, we will still be able to robustly constrain the upper and lower covering factor limits. We will illustrate such an application in section \ref{sec:discuss}.

\subsubsection{NIR, MIR and FIR emission as covering factor proxy}
\label{sec:nmf-type1}

In the previous section we have used the total torus luminosity as a covering factor proxy. We will now investigate the luminosities of hot (NIR, 1--5~$\umu$m), warm (MIR, 5--25~$\umu$m) and cold (FIR, 25--380~$\umu$m) dust as covering factor estimators. We confirmed that the results are not sensitive to adjustments of the wavelength range of these three bands (e.g. by changing the mid-far IR limit to several values between 20 and 100~$\umu$m).

\begin{figure*}
\centering
\includegraphics[height=0.49\textwidth]{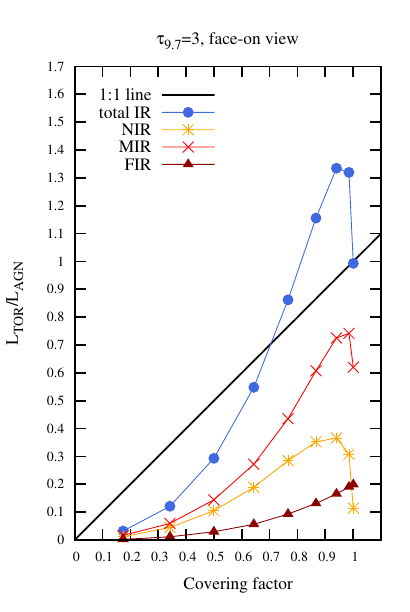}
\includegraphics[height=0.49\textwidth]{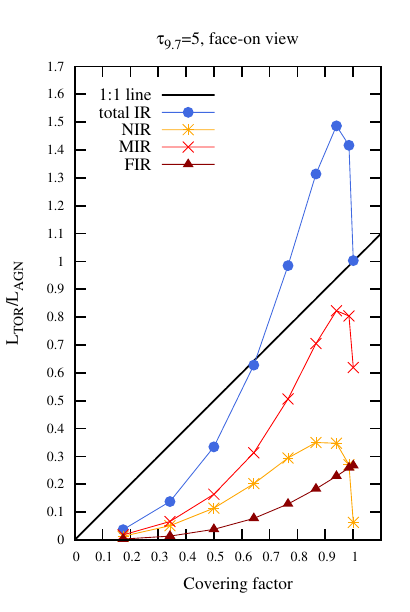}
\includegraphics[height=0.49\textwidth]{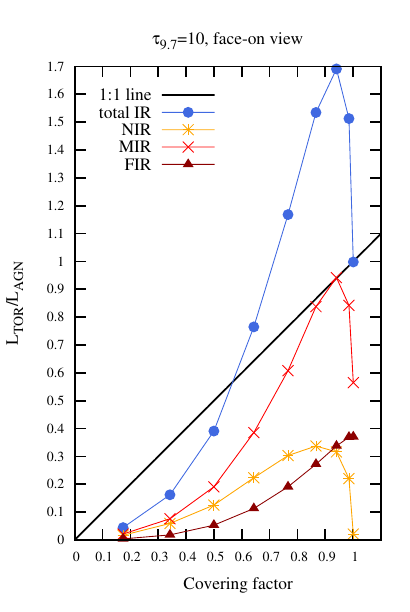}
\caption{Comparing $L_{\text{torus}}/L_{\text{AGN}}$ with covering factors, using the total torus IR (blue dots), NIR (yellow stars), MIR (red crosses) and FIR (brown triangles) luminosities and for $\tau_{9.7}=3, 5, 10$ in the left, central and right panels, respectively. Face-on (type 1 AGN) case.}
\label{fig:cfall}
\end{figure*}

In Fig.~{\ref{fig:cfall}} we show the same relation as in Fig.~{\ref{fig:cf-type1}} but for different wavelength bands, where $L_{\text{torus}}$ stands for total IR luminosity (blue circles; included for comparison), NIR luminosity (yellow stars), MIR luminosity (red crosses) and FIR luminosity (brown triangles). From left to right, panels correspond to a range of realistic cases of $\tau_{9.7}=3, 5, 10$. Since NIR-MIR-FIR luminosities represent a fraction of the total $L_{\text{torus}}$, they almost always underestimate the covering factor. We will now have a closer look at each of these.

By examining the yellow stars in all the three panels in Fig.~{\ref{fig:cfall}} we notice a favourable aspect: $L_{\text{NIR}}/L_{\text{AGN}}$ has almost no dependence on optical depth (and as we confirmed, neither on other parameters). This means that $L_{\text{NIR}}$ could be a very robust covering factor estimator. However, we must keep in mind that in the NIR band we can have significant contribution by host galaxy stellar emission and/or the accretion disk which must be carefully subtracted. Moreover, even when the stellar component is negligible or properly accounted for, the torus models are not always able to account for a significant fraction of the emission (the so-called NIR excess problem: \citealp{Polletta2008,Mor-Netzer-Elitzur2009,Deo2011,Mor-Trakhtenbrot2011,Lira2013}). The origin of this excess is unclear and plausible sources include hot graphite dust \citep{Mor-Netzer-Elitzur2009} and low density inter-clump dust in the two-phase torus model \citep{Stalevski2012a,Stalevski2012c}. Having in mind all this we advise extreme caution if $L_{\text{NIR}}/L_{\text{AGN}}$ is to be interpreted as the dust covering factor.

Now we turn our attention to the red crosses in Fig.~{\ref{fig:cfall}} which represent $L_{\text{MIR}}/L_{\text{AGN}}$. Again, the covering factors are almost always underestimated, except for highly obscured sources in the case of high optical depth (right panel). Similar as when using the total torus luminosity, the cases of different optical depths diverge significantly from each other only for high covering factors.

Finally, let us have a look at brown triangles in Fig.~{\ref{fig:cfall}} which stand for $L_{\text{FIR}}/L_{\text{AGN}}$. FIR emission mostly comes from the outer regions of the torus where cold dust resides, and from the dark side of the clumps, i.e. those which are not directly illuminated by the central source. But we must keep in mind that in the FIR, the host galaxy contribution is very often dominant over the AGN \citep[e.g.][]{Hatziminaoglou2008,Hatziminaoglou2009,Hatziminaoglou2010}. Disentangling the two is a challenging task and introduces more uncertainties. Hence, once again we advise extreme caution if $L_{\text{FIR}}/L_{\text{AGN}}$ is used to estimate the dust covering factor.

\subsection{$L_{\text{torus}}/L_{\text{AGN}}$ as a covering factor proxy: type 2 AGN case}
\label{sec:cf-type2}

In this section we examine the relation between $L_{\text{torus}}/L_{\text{AGN}}$ and covering factor for the type 2 AGN case, following the same approach as in Sec.~{\ref{sec:cf-type1}}.

\subsubsection{The total infrared emission}

Fig.~{\ref{fig:cf-type2}} is the equivalent of Fig.~{\ref{fig:cf-type1}}, except that now we are measuring the model SED luminosities for an edge-on view. We look first at the empty yellow diamonds in the left panel, which represent the case of an isotropic disk and a MIR thin torus. We note that this line is slightly more offset from the 1:1 identity line than in the type 1 case. Here we need to remember that MIR thin ($\tau_{9.7}=0.5$) refers to the underlying smooth dust density distribution, before the clumpy two-phase medium is generated (see Sec.~{\ref{sec:mod-par}}). The actual optical depth in the two-phase model may vary depending on the number of clumps along the particular line of sight, and even become optically thick, which is the reason for the offset in this case.

\begin{figure*}
\centering
\includegraphics[height=0.49\textwidth]{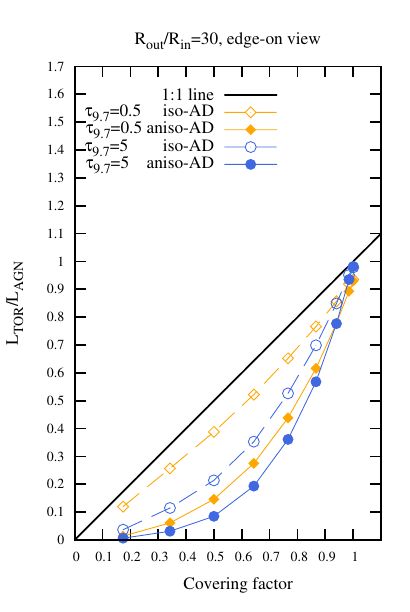}
\includegraphics[height=0.49\textwidth]{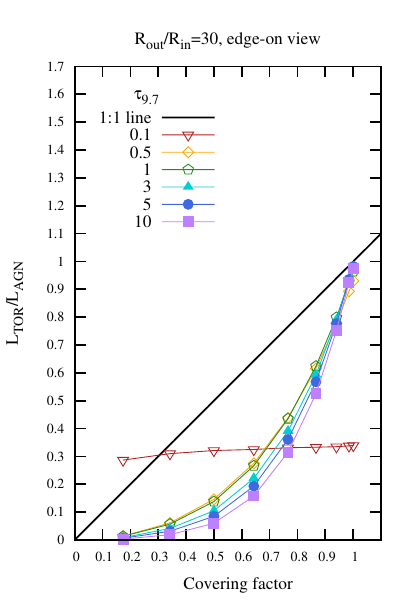}
\includegraphics[height=0.49\textwidth]{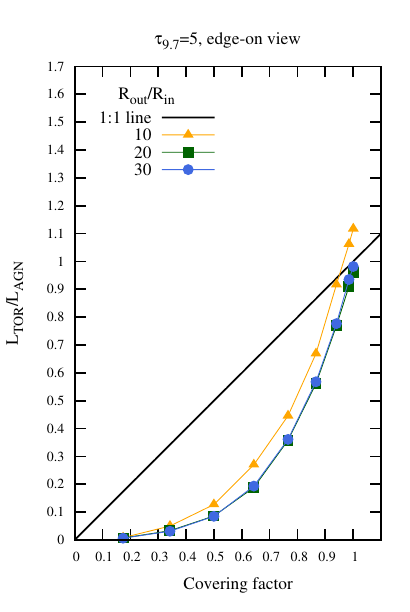}
\caption{The same as in Fig.~{\ref{fig:cf-type1}} but for the case of an edge-on view (type 2 AGN). We note that $L_{\text{torus}}/L_{\text{AGN}}$ always underestimates covering factors, with a weak dependence on the torus parameters.}
\label{fig:cf-type2}
\end{figure*}

Next we examine the empty blue circles, corresponding to the case of an isotropic disk surrounded by MIR thick dust. Since in the edge-on view the optically thick structure appears fainter than in the face-on view, using these fluxes we will underestimate $L_{\text{torus}}$ and thus the covering factor.

Moving on to the more realistic disk with anisotropic emission but encompassed by MIR thin dust (full yellow diamonds), we see that, as dictated by the disk emissivity dependence on polar angle (E.q.~(\ref{eqn:anisodisk})), the covering factors are severely underestimated.

Finally, in the most realistic case of a MIR thick dust enveloping an anisotropic accretion disk (full blue circles), based on previous considerations, we can only expect the covering factors to be even more underestimated; we see that this is indeed the case.

Now we turn to the central panel of Fig.~{\ref{fig:cf-type2}}, in which we examine the effect of changing the optical depth. Again, red empty triangles represent the case of very low optical depth, which means that the dust is partially transparent to the accretion disk emission, and thus $L_{\text{torus}}/L_{\text{AGN}}$ cannot be used as covering factor proxy. As for the other values of the optical depth we have considered, the relations do not diverge much from each other, especially if considering the realistic range of optical depths (full symbols). From the right panel, we see that there is very little dependence on the $R_{\text{out}}/R_{\text{in}}$ parameter.

We have so far focused just on models where the dust density does not vary as a function of the polar angle. If, instead, there is a gradient, we can expect some degeneracy with viewing angle (inclination). In Fig.~{\ref{fig:cf-type2-q1}} we inspect such a case with parameter $q=1$ as defined in Eq.~(\ref{eqn:dustdens}). We show only results for type 2 cases, leaving out the combinations of inclinations and covering factors that allow direct view of the central engine. For inclination in the range of $60-90^{\circ}$ (measured from the polar axis) the results are within the limits of optically thin to thick cases seen in Fig.~{\ref{fig:cf-type2}}. Only for tori with high covering factors ($>0.8$), which can provide obscuration for even small inclinations ($<50^{\circ}$), there is a significant departure between the $L_{\text{torus}}/L_{\text{AGN}}$--covering factor relations for different viewing angles. In the extreme cases, covering factors can be even somewhat overestimated.

\begin{figure}
\centering
\includegraphics[height=0.49\textwidth]{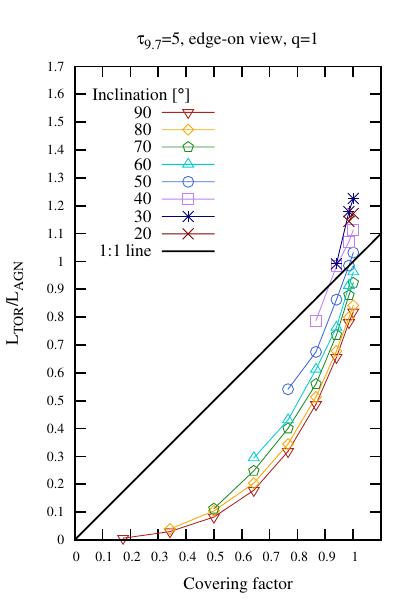}
\caption{Similar as in the central panel of Fig.~{\ref{fig:cf-type2}} but only for $\tau_{9.7}=5$ and $q=1$, i.e. with a gradient in dust density with polar angle (see Eq.~{\ref{eqn:dustdens}}). Only combinations of covering factors and viewing angles that have dust-intercepting lines of sight (i.e. type 2 AGN cases) are plotted. We see that even when the gradient is present, the resulting covering factors are still within the limits of optically thin to thick cases. The exception are tori with high covering factors ($>0.7$), which can provide obscuration for even small inclinations; in this case, covering factors could be overestimated.}
\label{fig:cf-type2-q1}
\end{figure}

\subsubsection{NIR, MIR and FIR emission as covering factor proxy}

We now inspect Fig.~{\ref{fig:cfall-i90}} where we can see how luminosities at different wavelength bands act as covering factor estimators. Like in Fig.~{\ref{fig:cfall}}, we consider NIR (1--5~$\umu$m, yellow stars), MIR (5--25~$\umu$m, red crosses), FIR (25--380~$\umu$m, brown triangles) and total IR (blue circles); from left to right, panels correspond to cases of $\tau_{9.7}=3, 5, 10$.

We see that with increasing optical depth, the NIR and FIR contributions are going in opposite ways: as contribution from hot dust is decreasing and becoming negligible, cold dust contribution rises and becomes comparable to that from warm (MIR) dust. The caveats mentioned in the Sec.~\ref{sec:nmf-type1} -- contamination by sources in the NIR and FIR other than AGN and the need for disentangling different components by spectral energy decomposition -- are to be kept in mind here as well . As such, neither NIR nor FIR luminosities are reliable covering factor proxies. On the other hand, in the case of type 2 AGNs, MIR luminosity is a much more robust covering factor estimator, as its contribution remains almost constant with increasing optical depth.

\begin{figure*}
\centering
\includegraphics[height=0.49\textwidth]{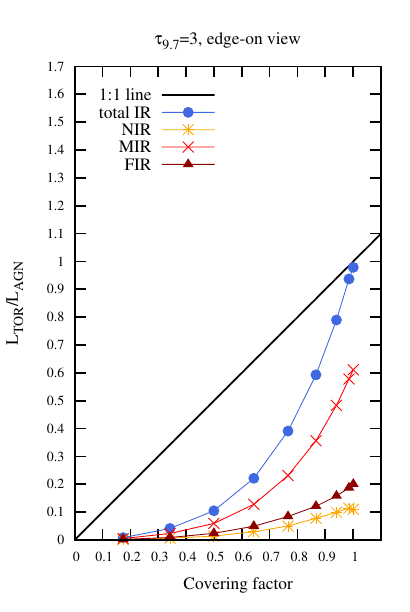}
\includegraphics[height=0.49\textwidth]{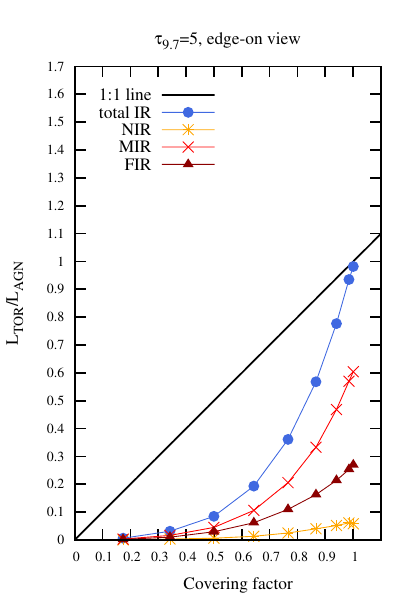}
\includegraphics[height=0.49\textwidth]{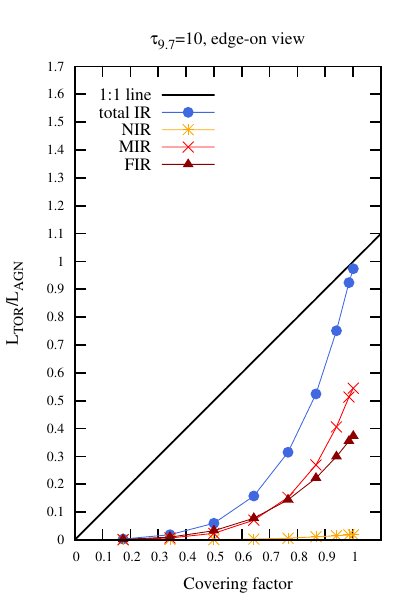}
\caption{The same as in Fig.~{\ref{fig:cfall}} but for the case of an edge-on view.}
\label{fig:cfall-i90}
\end{figure*}

\subsection{The case of misaligned accretion disk and dusty torus}

In all the results so far presented, we were assuming that the accretion disk and the dusty torus are aligned. This is a reasonable assumption if both the disk and the torus have an origin in the same accretion event, i.e. the same material arriving from larger scales to the vicinity of the black hole. An AGN can also be fueled by a series of such events coming in from random directions. When this happens a partial re-alignment of the newly arriving material and the old accretion disk must occur, producing a warped structure. \citet{Lawrence-Elvis2010} suggested that such a warped structure can provide the necessary covering factor. Detailed hydrodynamical simulations coupled with radiative transfer are needed to explore the evolution of such a structure. However, it is reasonable to assume that the accretion disk is able to partially clear out its close environment by dust sublimation, radiation pressure and winds, which would effectively reduce the misalignment with the dusty structure. In the remaining of this section, we will consider a case of a misaligned accretion disk and dusty obscuring structure, and study how this misalignment affects the covering factor estimates.

Unfortunately, precise estimates of inclination of the accretion disk and the dusty torus are available only for nearby sources that can be resolved at different wavelength regimes, and even then they may not be reliable. Usually, the position angles of the jet, ionization cone, maser disk and IR-emitting structure are used to estimate the orientation of the accretion disk and the dusty obscuring structure. The jet is assumed to be launched from the vicinity of the black hole, perpendicular to the disk. It is commonly believed that launching of the jet is related to the spin of the black hole, which is determined by the sum of angular momentums of all the accreting events in the past of an AGN. In that case, the presence of a strong and linear jet rules out multiple accreting events from random directions, as they would in a net effect reduce the spin of the black hole. In this scenario we expect that the disk-torus misalignment, if present, must be limited to rather small values. 

In the case of NGC 1068, there is a systematic change of orientation with radius. Summarizing the discussion from \citet{Lawrence-Elvis2010}, the inner radio jet is orthogonal to the polarization angle; the inner maser disk is misaligned with respect to the jet by $20^{\circ}$, while the outer maser disk is misaligned by $40^{\circ}$. The outer radio jet is aligned with the narrow line region and both are misaligned with the inner jet by $\approx 20^{\circ}$. Finally, \citet{Raban2009} conclude that the current phase of the AGN in NGC 1068 is not likely to have significant effect on the black hole spin (and thus the orientation of the jet), and that it is the angular momentum of the infalling material together with the gravitational potential of the nucleus, that are responsible for the current orientation of the maser, the accretion disk and the dusty torus. From similar considerations in the case of Circinus, a misalignment between the accretion disk and the dusty structure of $\approx 27^{\circ}$ is inferred \citep{Greenhill2003,Tristram2007,Tristram2014}. In a sample of 21 local objects, \cite{Asmus2016} studied the orientation between the resolved MIR elongated emission (tentatively interpreted as the inner funnel of an extended dust distribution) and the AGN axis inferred by different methods. They found that in 18 objects the MIR emission is aligned with the AGN axis within $35^{\circ}$, and in 9 of those within $18^{\circ}$; the median angular difference for the whole sample is $19^{\circ}$ with a standard deviation of $27^{\circ}$.

Having in mind all said above, we conclude that, while the estimates of the orientation between the accretion disk and the dusty structure remain uncertain and with a lot of caveats, the eventual misalignment is likely to be confined to up to $\approx 30^{\circ}$. 

To investigate effect of misalignments on the covering factor estimates, we carried out two additional sets of simulations, identical to those discussed previously, but with the accretion disk inclined by $15$ and $30^{\circ}$ with respect to the dusty torus. The resulting dust covering factor estimates in such systems are presented in Fig.~{\ref{fig:cf-misalign}}. Blue circles represent the previously discussed case of aligned disk and torus; yellow stars and red crosses stand for the cases of disk inclined by $15$ and $30^{\circ}$, respectively. Looking at the face-on case (left panel), we see that for $i_{disk}=15^{\circ}$ the change in estimated covering factors is marginal. For $i_{disk}=30^{\circ}$ the change is more significant. All the values are systematically higher than in the case of alignment; for lower covering factors, they almost lie on the 1:1 line. This is a consequence of the anisotropic disk emission: the more the disk is inclined, the more radiation will be reprocessed by the dust, resulting in higher IR emission, and thus higher covering factor estimates. In edge-on case (right panel) the misalignment does not play a significant role.

\begin{figure*}
\centering
\includegraphics[height=0.49\textwidth]{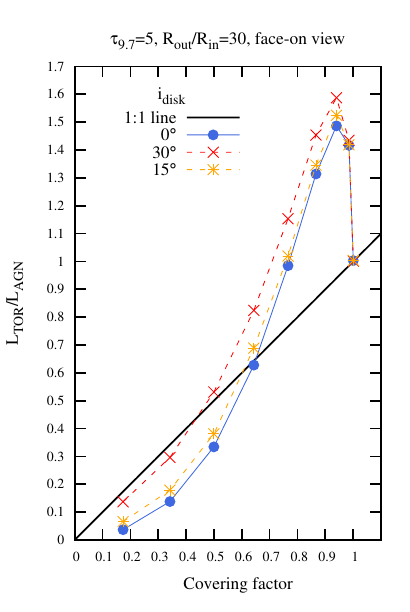}
\includegraphics[height=0.49\textwidth]{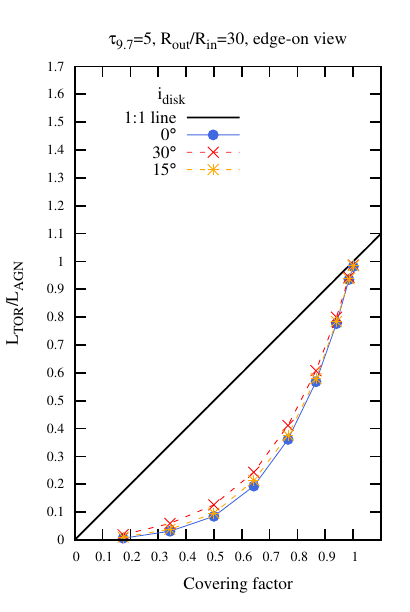}
\caption{Comparing the covering factor estimates for aligned (blue circles) and misaligned disk and torus cases. Disk tilted by $15^{\circ}$ (yellow stars) yields in a marginal difference. The difference becomes significant at a tilt of $30^{\circ}$ (red crosses).}
\label{fig:cf-misalign}
\end{figure*}

\section{Implications for inferring the obscured AGN fraction}
\label{sec:discuss}

\subsection{Obscured AGN fraction vs. $L_{\text{AGN}}$}

Now we will demonstrate a practical use of our results by correcting the observed $L_{\text{torus}}/L_{\text{AGN}}$ ratio to infer the obscured AGN fraction in two samples taken from the literature. We choose the works of \citet{Maiolino2007} and \citet{Lusso2013}, both dealing with type 1 AGNs only. The first work analyses a sample of 58 high-luminosity quasars and low-luminosity AGNs, in a redshift range $2<z<3.5$; the second one compiles a sample of 513 Type 1 AGNs from the XMM-COSMOS survey ($0.04<z<4.25$). We point the reader to the extensive comparison of the two samples and results in \citet{Lusso2013}.

\begin{table*}
\centering
\caption{Coefficients of the polynomial curve fits to the relation between the covering factor and $L_{\text{torus}}/L_{\text{AGN}}$ presented in the central panels of Figs.~{\ref{fig:cf-type1}} and {\ref{fig:cf-type2}}, and in Fig.~{\ref{fig:cf-misalign}} for a straightforward correction of observed luminosity ratio to account for the dusty torus anisotropy. The type 1 sections has an additional column indicating the maximum value of R for which polynomial approximation is valid. For type 2 corrections, $R_{max}$ is 1 in all cases.}
\label{tab:cf-lratio}
\begin{tabular}{|c|cccccc|c|ccccc|}
\hline
\multicolumn{13}{|c|}{${\text{CF}} = a_4R^4+a_3R^3+a_2R^2+a_1R+a_0$; $R\equiv L_{\text{torus}}/L_{\text{AGN}}$}                                                         \\ \hline
                      & \multicolumn{6}{c|}{\textbf{type 1}}              &  & \multicolumn{5}{c|}{\textbf{type 2}}              \\ \hline
\textbf{$\tau_{9.7}$} & $a_4$ & $a_3$ & $a_2$ & $a_1$ & $a_0$ & $R_{max}$ &  & $a_4$ & $a_3$ & $a_2$ & $a_1$ & $a_0$ \\
\textbf{3}            & -0.177798   & 0.875215   & -1.48727   & 1.40827    & 0.192478   & $1.3$   &  & 0   & 2.03866   & -3.97589   & 2.76458    & 0.204995   \\
\textbf{5}            & -0.0601471  & 0.47493    & -1.04546   & 1.20218    & 0.195615   & $1.5$   &  & 0    & 2.22844    & -4.28061   & 2.8593    & 0.22625   \\
\textbf{10}           & -0.0255416  & 0.299937   & -0.782418  & 1.02696    & 0.196387   & $1.7$   &  & 0   & 2.50138   & -4.68318   & 2.96875    & 0.258279   \\ \hline
\multicolumn{13}{|c|}{$\tau_{9.7}=5$; The disk and the torus misaligned}                                                          \\ \hline
\textbf{$i_{disk}$}   & $a_4$ & $a_3$ & $a_2$ & $a_1$ & $a_0$ & $R_{max}$ &  & $a_4$ & $a_3$ & $a_2$ & $a_1$ & $a_0$ \\
\textbf{15$^{\circ}$}            & -0.0845587 & 0.526976   & -1.08596   & 1.24635    & 0.151897   & $1.5$   &    & 0   & 2.08295   & -4.07339   & 2.80935    & 0.209374   \\
\textbf{30$^{\circ}$}            & -0.0341992 & 0.40201    & -1.06448   & 1.40097    & 0.003745   & $1.6$   &    & 0   & 1.94874   & -3.90322  & 2.81442    & 0.164062   \\ \hline
\end{tabular}
\end{table*}

Table \ref{tab:cf-lratio} contains coefficients of the polynomial curve fits,
\begin{equation}\label{eqn:polfit}
{\text{CF}}=a_4R^4+a_3R^3+a_2R^2+a_1R+a_0 ,
\end{equation}
to the relations between the covering factor (CF) and $R\equiv L_{\text{torus}}/L_{\text{AGN}}$ presented in Sec.~{\ref{sec:res}}, and in the central panels of Figs.~{\ref{fig:cf-type1}} and {\ref{fig:cf-type2}}. We provide the coefficients in the range of realistic values of $\tau_{9.7} = 3-10$. As explained in Sec.~\ref{sec:res}, the ${\text{CF}}-R$ curves in this range of optical depth values cover the most relevant parameter space, including variations of other torus parameters as well. The ${\text{CF}}-R$ relation was fitted only up to the turn-over point (${\text{CF}}=0.94$), because of the non-uniqueness of the relation beyond that point. As discussed in Sec.~{\ref{sec:cf-type1}}, this would have only a marginal effect on observational results, as there is an extremely small probability that AGNs with such high covering factors would be observed through the dust-free cone and classified as type 1s.

In Fig.~{\ref{fig:fobsc}} we show the obscured AGN fraction vs.\ $L_{\text{AGN}}$ from \citet{Maiolino2007} (left) and \citet{Lusso2013} (right), with solid black lines and points. The data we took are for the case the authors refer to as ``optically thin'', i.e., before applying the simple, analytical anisotropy correction factor (Eq.~3 in \citet{Lusso2013}). Both works found a decrease of the obscured AGN fraction with increasing $L_{\text{AGN}}$. Having a quick look back at Fig.~{\ref{fig:cf-type1}}, we remind ourselves that in the case of type 1 AGNs the dusty torus and the accretion disk anisotropies conspire to make $L_{\text{torus}}/L_{\text{AGN}}$ underestimate low covering factors, and overestimate high covering factors. The simple expression in Eq.~{\ref{eqn:polfit}} with coefficients in Table~{\ref{tab:cf-lratio}} allow us to straightforwardly correct the observed luminosity ratio. The coloured lines and symbols in Fig.~\ref{fig:fobsc} represent the obscured AGN fraction after we applied the corrections from the present work, for different optical depths (cyan, blue, violet) and disk-torus orientation (orange and red). We see that the actual trend of obscured fraction with $L_{\text{AGN}}$ is less steep than it originally appeared. We recover similar values as \citet{Maiolino2007} and \citet{Lusso2013} around $L_{\text{AGN}}\approx10^{45}$~erg/s, and have significant deviations at low and high $L_{\text{AGN}}$ ends. Note that there are considerable uncertainties in the bolometric correction factor for $L_{\text{AGN}}$ and dispersion of the obscured fraction around their median values. However, our approach allow us to put tighter limits on the obscured AGN fraction, compared to the limiting cases of ``optically thin and thick'' regimes discussed by the former studies (see Fig.~10a in \citealp{Lusso2013}). If we consider \citet{Lusso2013} results with our individual corrections, the dust covering factor of AGNs depend veary weakly on $L_{\text{AGN}}$, with values in the range $\text{CF} \approx 0.6-0.7$, depending on the employed correction. Note that the choice of the correction factors scales the results along the y-axis, while the slope remains almost unchanged.

\citet{Netzer2016} found that covering factors are basically independent of AGN luminosity and suggest that earlier results inferring steeper dependence of IR-derived covering factors on $L_{\text{AGN}}$ may be biased by the inconsistent use of various bolometric correction factors. The goal of the present work is to illustrate effects of the torus anisotropy and provide a method to properly correct for it; we do not attempt to revise the $L_{\text{AGN}}$ estimates. In the case Netzer et al. refer to as ``isotropic'' (before analytical anisotropy correction factor is applied), they found the median covering factor of $\text{CF} = 0.68$. As we have shown, for moderate optical depths in MIR, $L_{\text{torus}}/L_{\text{AGN}}$ is a very good indicator of the covering factors in the range of $\text{CF} \approx 0.65-0.7$. Thus, if we would apply our corrections to the obscured AGN fraction found by Netzer et al., their results would not change considerably. In fact, the very mild trends which might still be present within uncertainties, would become even flatter. However, Netzer et al.~sample is in a narrower luminosity range and at and almost without overlapping to those of \citet{Maiolino2007} and \citet{Lusso2013}. Given that the different methods are used to estimate $L_{\text{AGN}}$ in these works, further comparison between them would be precarious. It should be kept in mind that considerable observational uncertainties still remain, and changes of the covering factor could still be present and hidden within the uncertainties; but the change would be limited to less than a factor of 2 over more than 3 orders of magnitude in $L_{\text{AGN}}$.

\subsection{The discrepancy between the obscured AGN fraction inferred from IR and X-rays}

The decrease of the covering factor of the torus with increasing luminosity is systematically found in X-ray surveys \citep[e.g.][]{Ueda2003,Beckmann2009,Burlon2011,Ricci2013,Merloni2014,Ueda2014,Buchner2015}. In the most complete study of local AGNs, \citet{Ricci2015} reconstructed the intrinsic column density ($N_{\rm H}$) in two luminosity bins and found that the intrinsic fraction of both Compton-thin and Compton-thick objects decreases with $L_{\text{AGN}}$. Such trends may be interpreted as an indirect signature of AGN feedback, in a sense that more luminous sources are able to clear out their immediate environment (obscuring region) more efficiently through radiation pressure, winds, dust sublimation and photoionization (see e.g. the radiation-driven fountain scenario by \citealp{Wada2015}). However, there appears to be a discrepancy between the obscured AGN fraction found by X-ray studies and inferred from $L_{\text{torus}}/L_{\text{AGN}}$. Our revised results based on IR suggest: (a) obscured AGN fraction with very weak dependence on $L_{\text{AGN}}$, and (b) larger fraction of obscured high-luminosity AGNs than those estimated from X-ray surveys below 10 keV \citep{Hasinger2008}. 

The tension is reduced if we take into account the presence of AGNs with Compton-thick absorbers that are required in population synthesis models of the Cosmic X-ray background (CXB), but that are missed by surveys of individual sources \citep[e.g.][]{Ueda2014}. The two dashed lines in Fig.~{\ref{fig:fobsc}} (right) are the predictions based on the AGN population synthesis model by \citet{Ueda2014} calculated at $z=0$ (lower line) and $z\geq2$ (upper line). Here we convert the bolometric luminosity to the X-ray luminosity (2--10 keV) by using the luminosity-dependent bolometric correction factors given in \citet{Hopkins2007}. Then, we calculate the obscured fraction with $N_{\rm H} > 10^{22}$ cm$^{-2}$ including Compton-thick AGNs from the absorption function of \citet{Ueda2014}, which is dependent both on X-ray luminosity and redshift. We see that obscured AGN fractions at high luminosities ($\gtrapprox L_{\text{AGN}}10^{45}$~erg/s) with corrections for low and moderate optical depth fall within the region between the two boundaries. The tension remains at lower luminosities, but the redshift-luminosity coupling in the \citet{Lusso2013} sample must be taken into account for detailed analyses. The agreement with the CXB model predictions is better if a subsample excluding highly reddend AGNs from \citet{Lusso2013} is adopted. These highly reddend objects might have understimated $L_{\text{torus}}$ and thus, covering factors as well. In the subsample with $E(B-V)\leq0.1$, inferred covering factors are higher than in the total sample, by about 0.1. Note that the slope of the CXB model is much flatter than those found in X-ray surveys \citep[e.g.][]{Hasinger2008} because of (a) the inclusion of Compton-thick AGNs, whose number is the same as that of Compton-thin absorbed AGNs in the \citet{Ueda2014} model, and of (b) the non-linear relation between the bolometric and X-ray luminosities \citep[e.g.][]{Hopkins2007}. In addition, the positive correlation between luminosity and redshift in the sample would work to flatten the observed slope because higher obscured fractions are expected at higher redshifts (hence higher luminosities). 

Based on their IR-derived covering factors, \citet{Treister2008} also suggest a significant population of heavily obscured AGNs missed by X-ray observations. There are a number of other causes that could contribute to the discrepancy between IR and X-ray studies. \citet{Mayo-Lawrence2013} explore a scenario in which many objects are partially covered by Compton-thick material, and partially by intermediately thick material. Such objects would appear as Compton-thin, but with suppressed X-ray luminosity, artificially producing correlation of obscured fraction with luminosity. This does not pose a problem when data of sufficient quality is available, as then obscuration can be properly modeled. However, this is often not the case in large AGN surveys. \citet{Mayo-Lawrence2013} suggest that the observed trend can be reproduce in a model where 33\% of AGN are unobscured, 30\% are heavily buried, and 37\% have a range of intermediate partial coverings. On the other hand, if type 2 AGNs were systematically obscured by partially covering Compton-thick material, one would expect that the ratio between the Fe $K_{\text{alpha}}$ line and the 10-50 keV luminosity would be larger for type 2s than for type 1s, which is the contrary of what is observed \citep[e.g.][]{Ricci2014}. Other possible causes include a number of selection effects or unaccounted biases. \citet{Sazonov2015} showed that, apart from the negative bias in finding obscured AGN in hard X-ray flux limited surveys due to the absorption in the torus, there is also positive bias in finding unobscured AGN, due to Compton reflection in the torus. They demonstrate that these biases should inevitably lead to smaller obscured AGN fraction in the high luminosity end than in the low-luminosity end, even if the obscured AGN fraction has no intrinsic luminosity dependence. They also show that if the central X-ray source has a certain degree of anisotropy ($\sim cos\theta$), the intrinsic obscured AGN fraction could be consistent with a luminosity-independent torus covering factor. Further discussions on this and related issues can be found in \citet{Merloni2014} and \citet{Netzer2015}. 

Finally, the discrepancy could be a hint that the IR-emitting obscurer and the X-ray absorbing material may have different covering factors, owing to still poorly understood physics of dust and gas dynamics in a presence of a strong radiation field. Further work is needed to fully understand and resolve this issue.

\begin{figure*}
\centering
\includegraphics[height=0.49\textwidth]{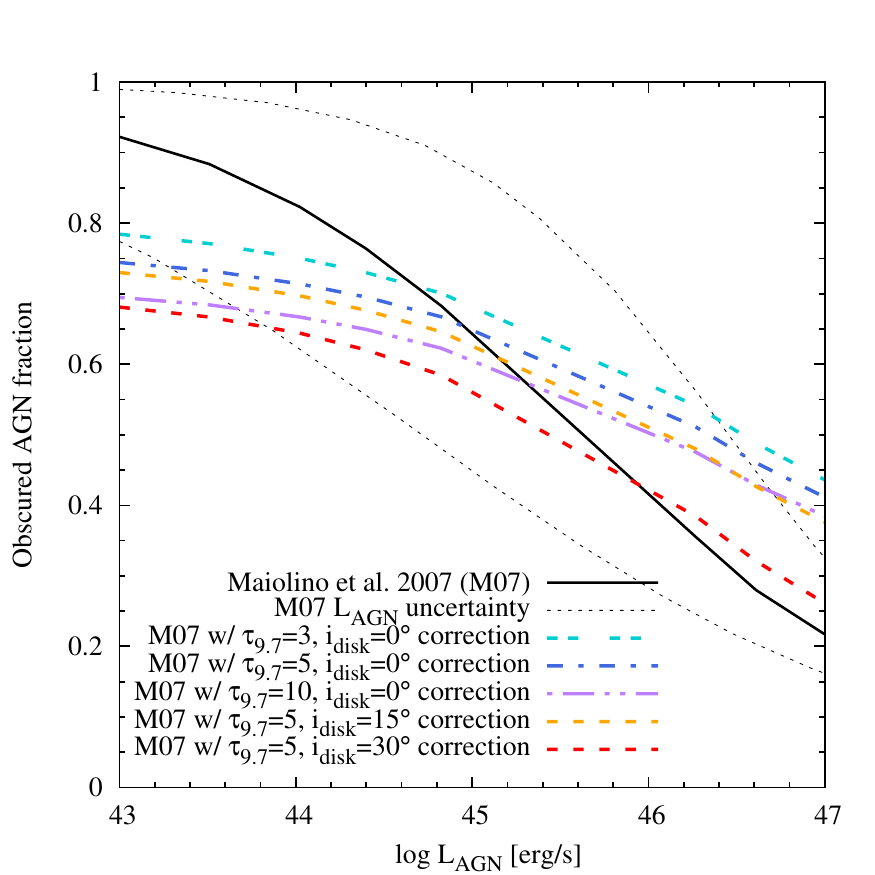}
\includegraphics[height=0.49\textwidth]{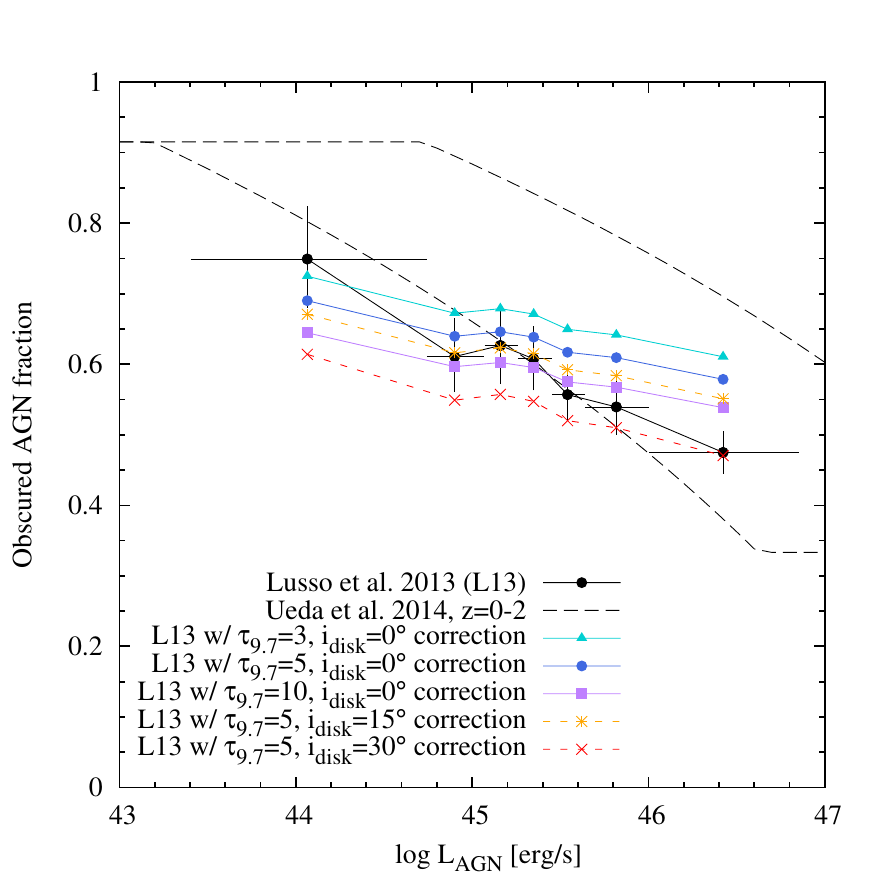}
\caption{The obscured AGN fraction vs.\ $L_{\text{AGN}}$ according to \citet{Maiolino2007} and \citet{Lusso2013} (left and right panel, respectively, in black solid lines and points), and the same after applying corrections based on our results (coloured lines and symbols). Cyan, blue and violet lines are for a range of realistic optical depths in the case of aligned disk and torus; yellow and red represent cases with misalignment. The dotted black lines in the left panel indicate the uncertainties due to the bolometric correction used in \citet{Maiolino2007}; uncertainty bars in the right panel represent $1\sigma$ of the distribution found by \citet{Lusso2013}. Uncertainties in the corrected cases are omitted for the clarity of the plot. Long-dashed black lines in the right panel are the prediction from the AGN population synthesis model by \citet{Ueda2014} calculated at $z=0$ (lower line) and $z\geq2$ (upper line). We notice that the trend of decreasing fraction of obscured AGNs with increasing luminosity becomes much less steep than in original data.}
\label{fig:fobsc}
\end{figure*}

\section{Conclusions}
\label{sec:con}

We conducted a comprehensive investigation of the $L_{\text{torus}}/L_{\text{AGN}}$ ratio as a dust covering factor estimator in AGNs. Using \textsc{SKIRT}, a state--of--the--art 3D Monte Carlo radiative transfer code, we have calculated a grid of SEDs emitted by the clumpy two-phase dusty structure surrounding the central engine (``the dusty torus''). With this grid of SEDs at hand, we studied the relation between the aforementioned luminosity ratio and the actual covering factor for different parameters of the torus. In this approach, anisotropies introduced by the nature of the accretion disk emission and by the complex interplay of different parameters during the transfer of radiation through the dust are taken into account. 

We found that the the combined effects of anisotropic emission of the dusty torus and the accretion disk lead to: 

\begin{itemize}
  \item $L_{\text{torus}}/L_{\text{AGN}}$ underestimating low covering factors and overestimating high covering factors in case of type 1 AGNs
  \item $L_{\text{torus}}/L_{\text{AGN}}$ always underestimating covering factors in case of type 2 AGNs.
\end{itemize}

Our results provide a new way to correct the observed $L_{\text{torus}}/L_{\text{AGN}}$ to account for the anisotropies and to recover the actual covering factors. We demonstrated the consequences of our findings for inferring the fraction of obscured AGNs, using the samples from the literature. 

In particular, our results suggest that:
\begin{itemize}
\item obscured AGN fraction depends more weakly on $L_{\text{AGN}}$ than previously thought, with covering factor values in range $\text{CF} \approx 0.6-0.7$ 
\item there is a larger fraction of obscured high-luminosity AGNs than that estimated from X-ray surveys.
\end{itemize}

We show that the larger fraction of obscured high-luminosity AGNs may be consistent with the presence of Compton-thick AGNs that are not included in analyses of absorbed AGN fraction using X-ray selected samples. The tension still remains at low-luminosities. Further investigation is warranted to fully resolve the discrepancy.

We emphasize that all the conclusions are consequence of just following two assumptions: (a) functional form of the accretion disk luminosity dependence on polar angle, and (b) the dusty structure surrounding the disk is of moderate- to high optical thickness in the MIR. The former assumption is certainly satisfied if the disk is geometrically thin. We showed that the second assumption must be satisfied as well for the typical dusty structures in AGNs, as otherwise the 10~$\umu$m silicate feature would be commonly appearing in emission (or at very least not in absorption) even in type 2 objects. Thus, as long as the two above mentioned conditions are met, our results hold up even if the exact geometry of the obscuring dusty material is different than assumed in this work (flared-disk).

Finally, we provide for the community polynomial function approximation of the relation between the covering factor and $L_{\text{torus}}/L_{\text{AGN}}$, which allows for a straightforward way to correct the observed luminosity ratio and obtain actual dust covering factors. We encourage colleagues to contact us if they should require correction functions for a specific sets of parameters.

\section*{Acknowledgments}

We thank the anonymous referee for useful comments and suggestions. MS is indebted to Peter Camps for continuous support and development of the \textsc{SKIRT} code and thankful to Daniel Asmus and Konrad Tristram for careful reading and comments on the manuscript. MS acknowledges support by FONDECYT through grant no.\ 3140518 and by the Ministry of Education, Science and Technological Development of the Republic of Serbia through the projects Astrophysical Spectroscopy of Extragalactic Objects (176001) and Gravitation and the Large Scale Structure of the Universe (176003). MS and CR acknowledge support from the Japanese Society for the promotion of Science (JSPS), through projects no.\ PE14042 and 12795, respectively. CR acknowledges financial support from CONICYT-Chile grants ``EMBIGGEN'' Anillo ACT1101, FONDECYT 1141218 and Basal-CATA PFB--06/2007. JF acknowledges the financial support from the grant UNAM-DGAPA-PAPIIT IA104015. Powered@NLHPC: this research was partially supported by the supercomputing infrastructure of the NLHPC (ECM-02).

\bibliographystyle{mnras}
\input{refs.bbl}

\appendix

\bsp

\label{lastpage}

\end{document}